\documentclass[letterpaper,amsmath,showpacs,amssymb,aps,nofootinbib,10pt,]{revtex4-1}
\usepackage[T1]{fontenc}
\usepackage[utf8]{inputenc}
\usepackage{amsfonts}
\usepackage{amssymb}
\usepackage{lmodern}
\usepackage{graphicx}
\usepackage{mathtools}
\usepackage[cmintegrals,cmbraces]{newtxmath}
\usepackage{verbatim}
\usepackage{ebgaramond}
\usepackage[dvipsnames]{xcolor}
\usepackage{geometry}
\usepackage{amsmath,amssymb,amsfonts,dcolumn,color,graphicx,graphics,latexsym,placeins,epsfig,tikz}
\usetikzlibrary{arrows,shapes}
\usepackage{subfigure,rotating,bm,mathrsfs}

\newcommand{\RN}[1]{
	\textup{\uppercase\expandafter{\romannumeral#1}}
}
\newcommand{\be}{\begin{equation}}
	\newcommand{\ee}{\end{equation}}
\newcommand{\ba}{\begin{eqnarray}}
	\newcommand{\ea}{\end{eqnarray}}

\usepackage[pagebackref=false, colorlinks=true]{hyperref}
\definecolor{redish}{rgb}{0.7,0.2,0.0}  
\definecolor{bluish}{rgb}{0.2,0.5,0.8}
\hypersetup{linkcolor=red,          
	citecolor=teal,        
	filecolor=magenta,      
	urlcolor=blue}          

\begin{document}

\title{ Time-dependent Hamiltonians and 
	Geometry of Operators Generated by Them}
\date{\today}

\author{Kunal Pal}  \email{kunalpal@snu.ac.kr} 
\affiliation{
	Department of Physics, Indian Institute of Technology Kanpur,  Kanpur 208016, India\\ and \\
	Department of Physics and Astronomy, Center for Theoretical Physics, Seoul National University, Seoul 08826, Republic of Korea}
\author{Kuntal Pal}\email{kuntalpal@gist.ac.kr}
\affiliation{
	Department of Physics, Indian Institute of Technology Kanpur,  Kanpur 208016, India\\and\\
	Department of Physics and Photon Science, Gwangju Institute of Science and Technology, 123 Cheomdan-gwagiro, Gwangju 61005, Republic of Korea}

\begin{abstract}
	We obtain the complexity geometry associated with the Hamiltonian of a quantum mechanical system, specifically in cases 
where the Hamiltonian is explicitly time-dependent. Using  Nielsen's geometric formulation of circuit complexity,  we calculate the bi-invariant cost 
associated with these  time-dependent Hamiltonians by suitably regularising their norms and obtain analytical expressions of the costs 
for several well-known time-dependent quantum mechanical systems. Specifically, we show that an equivalence exists between the total
costs of obtaining an operator through time evolution generated by a unit mass  harmonic oscillator whose frequency depends on time, and a harmonic 
oscillator whose both mass and frequency are functions of time. These results are illustrated  with several examples, including
a specific smooth quench protocol where the comparison of  time variation of the cost with other information theoretic quantities, such as the Shannon entropy, is  discussed.
\end{abstract}

	\maketitle
	\tableofcontents

\newpage

\section{Introduction}
The quantity computational complexity of a given quantum state roughly measures the minimum number of basic gates required to build the target state starting from an initial reference state. Among various measures to quantify computational complexity, one of the most commonly used is the one proposed by Nielsen and collaborators \cite{Nielsen1, Nielsen2, Nielsen3}. In this approach, the reference and target states correspond to two points on a curved manifold, and all the possible unitary transformations that connect these two points are assigned with a cost function. According to this geometric formulation, the path that minimises the cost function is nothing but the geodesic on that manifold of the unitary operator, and the length of the geodesic is called the Nielsen complexity (NC).  This particular measure of complexity has been explored in significant detail in recent literature not only because it can shed light on fundamental aspects of the information contained in a quantum state \cite{Myers1, BSS, KKS, Myers2, HM, BMS, BS, Caceres:2019pgf}, but also due to the fact that NC can also be used an important probe of quantum quenches \cite{ABHKM, Gautam:2022gci, Pal:2022rqq, Chandran:2022vrw, Camargo:2018eof}, quantum phase transitions \cite{LWCLTYGG, JGS, XYZ, SK, JGS2, PPS, WH, Roca-Jerat:2023mjs, Jaiswal:2024yts, Jayarama:2022xta}, and quantum chaos \cite{BDKP1, BDKP2, HJU, Qu:2022zwq}, 
and has various  other  applications as well  \cite{Pal:2022ptv, Chowdhury:2023iwg, Adhikari:2021ckk, Basteiro:2021ene, Camilo:2020gdf, Erdmenger:2020sup}.  
Specifically, typical behaviour of circuit complexity has been used in \cite{ABHKM, Camargo:2018eof, Pal:2022rqq, Gautam:2022gci, Chandran:2022vrw} to 
understand  the unitary dynamics of an isolated quantum system, when the system is taken out of equilibrium by means of a single or multiple sudden or smooth quenches thereby changing the parameters of the Hamiltonian according to different relevant time scales. It has also been used in various works as an important quantum information theoretic tool to detect the zero temperature quantum phase transition in various contexts and models, where typically the discontinuity of the complexity marks the quantum critical point  \cite{LWCLTYGG, JGS, XYZ, SK, JGS2, PPS, WH, Roca-Jerat:2023mjs, Jaiswal:2024yts, Jayarama:2022xta}.  Apart from the NC, there are also other important measures of circuit complexity of quantum states 
which  differ from Nielsen's approach in many aspects (e.g., the one introduced in \cite{CHMP} by using the Fubini-Study metric associated with a pure quantum state), 
however, in this paper, we shall 
mostly consider Nielsen's formulation.  For a recent review of different approaches to circuit complexity and its application in holography, we refer to \cite{Chapman:2021jbh}.

It is important to note that, apart from the given target and reference states,  the NC  crucially depends on the choice of the cost function associated with each of the unitary paths on the manifold. Though  the qualitative nature of NC in a quantum system strongly depends on the choice cost function,
in Nielsen's  original formulation, this choice remains arbitrary (for an exposition on different cost functions used in this context, we refer to \cite{BMS}).
Another built-in aspect of NC is the fact that the unitary complexity geometry associated with NC is either right or left-invariant under a global transformation of the generator of the unitary curve connecting initial and final operators.  As was pointed out in \cite{YK, YK2, YANZK, YANZK2}, for a physical quantum system, 
	when this generator is the Hamiltonian of the system itself, the lack of unitary invariance is somewhat unsatisfactory since, e.g.,  the Nielsen complexity geometry corresponding to two Hamiltonians related by a unitary transformation would be different,
	and hence unacceptable from a physical point of view since they describe the same quantum system.  Based on the assumption that NC and hence 
	the cost function should be invariant under both left and right unitary transformation of the generator associated with a given path on the unitary manifold, the  notion  of bi-invariant complexity was proposed in \cite{YK, YANZK, YANZK2}. In this paper, we mainly consider this bi-invariant formulation of the 
	NC,  and apply it to the Hamiltonians of some well-known quantum mechanical systems. 

Usually, in the problem of calculating NC (or other notions of circuit complexity), the following situation is considered: the reference and target quantum states are  specified (along with the chosen cost functional and penalty factors for any gate directions, if any), and one has to minimise the costs
associated with all the relevant unitary curves  to find the NC. However, in this approach the instantaneous generator of the unitary curve is not usually given specific importance since one is only interested in the length of the geodesic curve (i.e., complexity) between two states (or operators) on the unitary complexity geometry (CG), irrespective of the generator of that path. 
Consider now a different posing of the above-mentioned problem, which is equally important: the time evolution of an  initial reference under the Hamiltonian of a given quantum system.  In that case,  the unitary operator is 
the time evolution operator itself, and the target state is actually some time-evolved state for a particular value of time.  
Of course, any arbitrary target state may not be on the unitary curve generated by the system Hamiltonian, and consequently, those states are not achievable under this protocol.

From a different perspective,  suppose that one can choose only two from the following  three quantities: a reference state, a target state, and 
	the (Hermitian) generator of the path connecting  these two states  on a CG, while the third one is 
	naturally constrained by the unitary evolution. In the standard way of defining NC, the first two   (i.e., reference and  target states) are given, whereas here we are interested in the 
	scenario where the first and third options (i.e., reference state and system Hamiltonian) are pre-assigned. As a natural consequence of the preceding statement, the nature of the parameter used to parametrise the unitary curve connecting the reference and target state is different from the usual treatment of NC.  
	Specifically, in this paper, we construct a bi-invariant CG associated with the operators generated by the Hamiltonian time evolution of quantum mechanical systems. 
	Therefore, in that case, the parameter along the unitary path connecting the reference and target state is  time. And the reason
	for choosing the bi-invariant geometry for this purpose is the fact that, on the one hand, it is invariant under unitary transformations which leave the Hamiltonian invariant, while on the other hand, since this geometry solely depends on the generator of the unitary evolution curve, i.e., here the Hamiltonian of the quantum 
	system under consideration, its computation is relatively straightforward for simple systems. In summary, our goal 
	here is to calculate the cost of operators along the time evolution curve of a quantum mechanical system by using the framework of  bi-invariant CG.

The quantum mechanical  systems  we mainly focus on have Hamiltonians which are explicit functions of time, and  their associated Hilbert space is infinite-dimensional.  These two characteristics (explicit time dependence and infinite-dimensional Hilbert space) of the Hamiltonian have important bearings 
on the cost associated with time-evolution operators.  Firstly, due to the fact the generator (here the Hamiltonian) is an explicit function of the parameter (the time), and 
since the CG is bi-invariant, the unitary curve is not a geodesic (note that the geodesic curves on a bi-invariant geometry are determined by 
generators which are constant along these trajectories).
Therefore, in this set-up it is only meaningful to consider the total cost along the given curve, not the minimum among a set of such costs computed along a 
collection of trajectories connecting a given target and reference state.   Secondly, due to the infinite-dimensional Hilbert space associated with the generator of the
unitary curve, to obtain a meaningful expression of the cost we need to suitably regularise it.  Here we roughly follow the method outlined in \cite{YK}
for the case of  time-independent Hamiltonians and suitably generalise it for cases when the generator is explicitly  time-dependent.

\textbf{Summary of the paper. } We now summarise the content and main results obtained in the rest of the paper. 
In the next section, we shall briefly review the geometric formulation of the circuit complexity pioneered  by Nielsen. Also, we shall introduce
the bi-invariant CG and the cost that will be used in the rest of the paper.   In section \ref{time-independent} we use this cost function
to obtain the CG associated with time-independent Hamiltonians, where we also discuss the procedure of regularising these costs. 
In that section we also show how the addition of an extra linear force term in the Hamiltonian can be interpreted as a trace regulator, thereby providing a physical 
justification for such method.  Since for the examples considered in this section, the Hamiltonian is time-independent, the corresponding 
time-evolution is actually a geodesic in the bi-invariant geometry, and hence, the cost computed along such a trajectory is the bi-invariant complexity itself. 

Section \ref{OTFCG} deals with the computation of the bi-invariant CG associated with a harmonic oscillator (HO) Hamiltonian whose frequency depends on time.  
One of the main ingredients of our analysis is the eigenfunctions of the time-dependent invariant operator which arises in a formulation developed by Lewis and 
Riesenfeld  \cite{LR, DR}. By using the matrix elements of the Hamiltonian in such a wavefunction basis, we obtain a compact analytical formula for the regularised 
cost function  (see eq. \eqref{NormH5final}), which constitutes one of the main results of this paper. We also show how this time-dependent cost reduces to the more familiar expression for the cost function of a
usual HO  in the appropriate limit and obtain a bound on the total cost evaluated along the 
time-evolution curve up to a certain finite time (see eq. \eqref{bound}). 

In Section \ref{OTMFCG}, we generalise the results of section \ref{OTFCG} for the case of the Hamiltonian
of an oscillator whose both mass and frequency  changes with time. For this Hamiltonian also, we are able to provide an analytical formula for the regularised 
cost function by using the Lewis - Riesenfeld theory of time-dependent invariant operators. The fact that such a Hamiltonian is actually an element 
of the $su(1,1)$ Lie algebra provides a way of directly obtaining the exact form of the time evolution operator using techniques of Lie group theory.
Using this exact expression for the time-evolution operator, we show that suitably redefining the time parameter, it is possible to map the  time evolution 
operator of an oscillator with time varying mass and frequency to an oscillator with unit mass but time varying frequency. Furthermore, we show that 
under the same reparametrisation, the total bi-invariant cost evaluated along the time-evolution curves of these two oscillators can be mapped to each other. 
In section \ref{quench} we compute the bi-invariant cost of a oscillator with time-dependent frequency and compare it with other measures, 
such as the change in Shannon 
entropy in one of the most common protocols where such time-dependent Hamiltonians arise, namely that of a smooth quantum quench. 

We discuss the significance of our results in section \ref{conclusion}, and indicate a few possible future directions. This paper also 
contains three appendices; in Appendix. \ref{timeindexamples} and \ref{ckosc}, we discuss further examples of calculation of the bi-invariant 
geometry of, respectively, time-independent and explicit time-dependent Hamiltonians, whereas in Appendix. \ref{decomp}
we provide some details of the computation of the time-evolution operator associated with Hamiltonians which are elements of the $su(1,1)$ Lie algebra. 


\section{Cost functional and the circuit complexity in Nielsen's approach }\label{costNC}
In this section we briefly review Nielsen's geometric approach of calculating the circuit complexity \cite{Nielsen1,Nielsen2,Nielsen3}. 
Suppose, for a quantum system we are given  a set of unitary operators, and we want to prepare a state $\big|\psi_{T}\big>$, known  as the target state, starting from a given reference state  $\big|\psi_{R}\big>$.  In the problem of calculating the circuit complexity we consider the following question.  What is the minimum number of operators within this set that we require  to accomplish this transformation? This  is known as the computational complexity of preparing  the state $\big|\psi_{T}\big>$ from $\big|\psi_{R}\big>$ denoted as $\mathcal{C}\Big(\big|\psi_{R}\big> \rightarrow \big|\psi_{T}\big>\Big)$.  

In Nielsen's approach, one considers a state 
infinitesimally different from the reference state such that a unitary operator can produce the former starting from the latter with an infinite accuracy. We denote the unitary operator connecting the reference and the target state to be $O$, i.e. $\big|\psi_{T}\big>=O\big|\psi_{R}\big>$. Nielsen's approach is to consider a space of such unitary operators on which the identity operator and the required operator $O$  can be thought to be  connected by a continuous curve $c(s)$, parameterised by some affine  parameter $s$. Thus, each point on such a curve represents a unitary operator $c(s)$, and we parameterise the path in such a way that  its starting point (the identity operator) is at $s=0$ and the final point (the operator $O$) corresponds to $s=1$. Hence,  $c(s=0)=\mathbf{I}$ and $c(s=1)=O$.

A quantity important for our purposes is the so-called instantaneous Hamiltonian $H(s)$ \cite{BMS},  which is nothing but the generator of the infinitesimal translation along the curve $c(s)$ and hence, satisfies the usual relation 
\begin{equation}\label{Hamiltonian}
i\frac{dc(s)}{ds}=H(s)c(s)~.
\end{equation}
Given an arbitrary state  $\big|\psi(s)\big>$,  the action on it by $H(s)$  takes it to a nearby state ($\big|\psi(s+ds)\big>$) 
 infinitesimally different from it. Therefore, the instantaneous state $\big|\psi(s)\big>$ satisfies  following the Schrodinger like equation
\begin{equation}\label{Schrodinger}
i\frac{d\big|\psi(s)\big>}{ds}=H(s)\big|\psi(s)\big>~.
\end{equation}
For quantum systems, this equation is the same as the Schrodinger equation  where the system Hamiltonian explicitly depends on time $t$
(in that case, the parameter we call $s$ is just the time) and equivalently, Eq. (\ref{Hamiltonian}) is the same as that is satisfied by  the time evolution operator $\mathcal{U}(t)$. This analogy with the time evolution operator  directly indicates a way of writing the operator $c(s)$ as a path-ordered integral over the Hamiltonian $H(s)$ as 
\begin{equation}\label{pathordered}
c(s)=\overleftarrow{\mathcal{P}}\exp\bigg[-i\int_{0	}^{s}H(s^{\prime})\text{d}s^{\prime}\bigg]~.
\end{equation}
This identification  is very useful for our purposes since it enables us to calculate the NC of
such time evolution  operators generated by the Hamiltonians of the time-dependent quantum systems.

There are infinitely many different continuous paths connecting the identity with the desired operator $O$, and in Nielsen's approach, 
the next step is 
to associate a certain computational cost with each of them by  defining a cost function $F\big(c(s),~H(s)\big)$,
which can, in general, depend on both the unitary curve and the generator.  Then, the NC of creating the operator $O$ starting from the identity  is obtained by minimising the cost functional (which is the integral of the cost function over a specific curve
$c(s)$). Mathematically, to find out the NC of an operator $O$ starting from the identity operator we need to solve the following minimisation problem,
\begin{equation}\label{complexitydef}
\mathcal{C}\big(O\big)=\min \int_{0	}^{1}F\big(c(s),~H(s)\big) \text{d}s~~~\text{with}~~c(s=0)=\mathbf{I}~,~\text{and}~~c(s=1)=O~.
\end{equation}
Now there can be many such operators $O_{i}$, belonging to the general set $O$, which connects the final state $\big|\psi_{T}\big>$ with the initial state  $\big|\psi_{R}\big>$. The NC of preparing $\big|\psi_{T}\big>$ from $\big|\psi_{R}\big>$ is the minimum value of the complexity of all such operators \cite{BS} i.e.
\begin{equation}
\mathcal{C}\Big(\big|\psi_{R}\big> \rightarrow \big|\psi_{T}\big>\Big)=\min \Big\{\mathcal{C}(O_{i})~ |~ \forall O_{i}~\in O~,~\big|\psi_{T}\big>=O\big|\psi_{R}\big>~~\Big\}~.
\end{equation} 
In this paper we mainly consider the NC and its  associated geometry (to be described below) of those operators for which the generator in Eq. (\ref{pathordered}) is that of the Hamiltonian of some quantum system.

Though in Nielsen's original formalism there are no physical principles by which these cost functions can be fixed, nevertheless, one can  choose the cost function in such a way that it satisfies some generic properties such as smoothness, continuity, positivity, and positive homogeneity. Furthermore,  $F$ can be chosen in such a way that it satisfies the triangle inequality for the sum of two Hamiltonians (see \cite{Nielsen1, Myers1} for a detailed discussion of these properties of the cost). When the cost 
is chosen so that it obeys these requirements  the complexity in Eq. (\ref{complexitydef}) actually  represents the length of a geodesics on a Finsler manifold. However,
even after imposing all these requirements on the cost, it is not unique.  In literature there are many choices for these cost functions, which are sometimes chosen in such a way that the final expression for the  NC comes with some required scaling behaviours, see, e.g., \cite{Myers1,Myers2,CHMP}. For a summary and relative merits and demerits  of different instantaneous state-independent and state-dependent cost functions used in the literature, see \cite{BMS}.

As mentioned above, since  the unitary operators we shall  consider in this paper are generated by the Hamiltonian of some quantum system, we assume that the Hamiltonian can be written in terms of a basis $\{e_{I}\}$, with respective coefficients denoted as $Y^{I}$, i.e. $H
=e_{I}Y^{I}$. In terms of these coefficients, two popular state-independent cost functions  used widely in the literature are 
\begin{equation}
F_{2}=\Big[\sum_{I}\Big(Y^{I}(s)\Big)^{2}\Big]^{1/2}~~,~~\text{and}~~F_{\kappa}=\sum_{I}\big|Y^{I}(s)\big|^{\kappa}~~.
\end{equation}
The first one of the above cost functions, as well as the second one with $\kappa=1$ were originally proposed by Nielsen, while the second class of costs with 
$\kappa>1$ were proposed in \cite{Myers1}. As pointed out in \cite{BMS}, there is a problematic issue with all these costs, namely, unless one imposes certain extra criteria on the bases used in expanding the instantaneous Hamiltonian, the geometry  resulting from two different choices of this basis do not in general coincide.  The simplest such condition one can impose is $\text{Tr}\big[e_{I}.e_{J}\big]=N\delta_{IJ}$, $N$ being an overall normalization constant.

The construction described above naturally represents a geometry in the space of unitaries, known as the CG.
To see this,  we write down an element of the unitary curve  in the following suggestive way (see Eq. (\ref{complexitydef}))
\begin{equation}
\text{d}l^{2}= F^{2}\big(c(s),H(s)\big)\text{d}s^{2}=g\big(H(s),H(s)\big)\text{d}s^{2}~.
\end{equation}
Written in terms of the basis $\{e_{I}\}$, defined above we then have the metric components of the CG to be 
\begin{equation}\label{Compgeometry}
F^{2}\big(H(s)\big)=g_{IJ}Y^{I}Y^{J}~.
\end{equation}
This formula clearly indicates that the cost function actually gives norm of the instantaneous Hamiltonian \cite{YK}. Therefore, below, we sometimes use these two words 
interchangeably.

Now we go back to the problem of defining such a geometry uniquely. Besides the above mentioned problem of nonuniqueness of the
CG, there is one additional feature of such constructions - the  resulting geometry is only right (or left) 
invariant under a global transformation of the unitary operator. As discussed in detail in \cite{YANZK,YANZK2,YK}, this property is somewhat
problematic for defining complexity associated with a quantum mechanical Hamiltonian, since a unitary transformed system  
is identical to that  of the original system, in the sense that physical observables computed from both of them make 
identical predictions for the original system and the transformed one.   Hence, it is very much desirable that for an arbitrary unitary operator $U(s)$,  defined on the CG,  the cost function or the norm should be such that it is invariant 
under the unitary transformation of the Hamiltonian, i.e., 
\begin{equation}
F\big(H(s)\big)=F\big(U(s)H(s)U^{\dagger}(s)\big)~,
\end{equation}
and as was shown in \cite{YANZK}, this property, in turn, guarantees that the resulting NC is invariant under such a transformation. 
This property of the CG is called  bi-invariance. By requiring that CG associated with  a  quantum system be bi-invariant, the following  class of cost functions 
has been obtained  in \cite{YANZK,YANZK2}
\begin{equation}\label{pcosts}
F\big(H(s)\big)=\lambda_{0}\bigg[\text{Tr}\Big[\big(H(s)H^{\dagger}(s)\big)^{p/2}\Big]\bigg]^{1/p}~,
\end{equation}
where the constant  $p$ takes positive integer values, and $\lambda_{0}>0$ is an overall constant multiplier. The resulting cost function is  known as 
the Schatten norm.

In this paper we shall  consider the above norm with $p=2$, since as argued in \cite{YANZK2}, the expression for cost function (and the resulting expression for the NC) 
for different values of $p$ are quantitatively the same,  only their numerical values are different.
The aim of this paper is to calculate the cost of operators generated by the Hamiltonian of a quantum system, both time-independent and explicitly time-dependent cases, with a specific focus given to the latter case due to its more involved nature.

\section{Complexity geometry associated with time-independent Hamiltonians} \label{time-independent}
We now discuss the construction of the bi-invariant CG  of operators generated by the Hamiltonian of some well-known quantum systems. 
In this section we consider the cases where the Hamiltonian is time-independent, while in the next section, and the rest of the paper, we deal with
systems whose Hamiltonians are explicit functions of time.  The instantaneous Hamiltonian appearing in Eq. (\ref{Schrodinger}) is the ordered Hermitian Hamiltonian operator (which we denote as $\mathbf{H}$ in the followings) of a given quantum system, and this equation is nothing but 
the Schrodinger equation for the time evolution of the state $\big|\psi(t)\big>$. Furthermore, the unitary operator of our concern 
is the time evolution operator generated by the system Hamiltonian.  At this point, we stress that  we want to evaluate  
the computational cost along the time evolution curve $\mathcal{U}(t)$, not the NC of a time evolved state (or an operator on the time evolution curve)
-  this fact  makes our analysis different from other works where  time evolution of NC  corresponding to different quantum systems have been reported \cite{BDKP1,YK2,HJU}. Specifically, in our set-up,  the parameter $s$ characterising the curves in the space of unitary operators is time $t$. \footnote{In this sense, the motivation of the present paper is similar to the recent set of 
	works \cite{Craps:2022ese, Craps:2023ivc, Craps:2023rur}, which has obtained an upper bound on the NC by restricting the minimisation procedure described in the previous section within a prescribed 
	family of curves connecting the identity and the target operator. }


When $p=2$, with the Hamiltonian operator as the generator, the cost functions in Eq. (\ref{pcosts}) reduce to 
\footnote{From this section onwards, the instantaneous Hamiltonian is the Hamiltonian of a quantum system,  which we denote as $\mathbf{H}$. 
Most such quantum mechanical operators will be represented by boldfaced letters. }
\begin{equation}\label{p=1cost}
F^{2}\big(\mathbf{H}(s)\big)=\lambda_{0}^{2}\text{Tr}\big[\mathbf{H}^{2}(s)\big]~.
\end{equation}
In that case, the bi-invariant cost represents a length on a  Riemannian metric, which is nothing but  the Cartan-Killing metric corresponding 
to  the Lie group whose generator 
is the Hamiltonian of the quantum system under consideration \cite{RG,FSS}. For the systems we consider in this paper, the Hamiltonian is an element of some Lie algebra, and the above mentioned Lie group is associated with this algebra. Clearly,  to find out the CG using this cost, we need to calculate the trace of the squared Hamiltonian. However, the  Hilbert spaces associated with the systems we are going to consider are infinite dimensional, and hence the trace can be divergent.
A procedure of obtaining  the CG by suitably regularizing the trace of the Hamiltonian operator using the Hurwitz $\zeta$ functions was described in \cite{YK}. For later purposes, we first briefly review this procedure  and its application to find out the   CG of a generalised oscillator, as worked out in \cite{YK}. We can also provide an alternative interpretation of this regularization in terms of a constant driving force present in the original Hamiltonian (see subsection \ref{driving}). 

The first step is to calculate  the  ``mean value'' of the Hamiltonian $\mathbf{H}$ (denoted here by $a$), by  imposing the following condition on the trace 
\footnote{Since in this section we only consider systems whose Hamiltonians are not explicit functions of time, the mean value $a$ is a constant. 
	We shall consider more general time-dependent systems in the next section where this quantity  will also be time-dependent. }
\begin{equation}\label{regularization}
\text{Tr}\big[\mathbf{H}-a\mathbf{I}\big]=0~.
\end{equation} 
We refer to this condition as the \textit{regularization condition} in the following. Since this is an $U(1)$ transformation, the addition by $-a\mathbf{I}$ does  not alter the CG, i.e. the bi-invariant geometry generated by the original Hamiltonian $\mathbf{H}$ 
is the same as the one generated by the  \textit{displaced Hamiltonian} $\tilde{\mathbf{H}}=\mathbf{H}-a\mathbf{I}$ \cite{YK}. 
With $a$ determined from the above condition, the regularised value of the cost  in Eq. (\ref{p=1cost}) is given by the trace of the  squared displaced Hamiltonian  (the subscript $r$ implies regularised)
\begin{equation}\label{regularisedcost}
F^{2}_{r}\big(\mathbf{H}\big)=\lambda_{0}^{2}\text{Tr}\big[\big(\mathbf{H}-a\mathbf{I}\big)^{2}\big]~.
\end{equation}
To evaluate this quantity, we first denote the diagonal matrix elements of $\mathbf{H}$  with respect to a complete set of orthonormal eigenfunctions $\big|\psi_{n}\big>$, which diagonalise the Hamiltonian, as $E_{nn}$. These are just the energy eigenvalues for time-independent Hamiltonians considered in this section. 
Then, one defines the analytic continuation of the sum 
\begin{equation}
\zeta_{\mathbf{H}}(k)=\sum_{n=0}^{\infty}\frac{1}{\big(E_{nn}\big)^{k}}~
\end{equation}
as the  $\zeta$ function associated with the Hamiltonian. Now it is easy to see that the  trace of the Hamiltonian  is just $\zeta_{\mathbf{H}}(-1)$. Similarly, for the displaced Hamiltonian $\tilde{\mathbf{H}}=\mathbf{H}-a\mathbf{I}$, the associated $\zeta$ function can be written as 
\begin{equation}\label{zetaHdis}
\zeta_{\tilde{\mathbf{H}}}(k)=\sum_{n=0}^{\infty}\frac{1}{\big(E_{nn}-a\big)^{k}}~~.
\end{equation}
Hence,  the regularised trace of the Hamiltonian is given by $\zeta_{\tilde{\mathbf{H}}}(-1)$. The last step of finding out the regularised cost is to recognise that it can be written in terms of the above  zeta function as 
\begin{equation}
F^{2}_{r}\big(\mathbf{H}\big)=\lambda_{0}^{2}\text{Tr}_{r}\big[\big(\mathbf{H}-a\mathbf{I}\big)^{2}\big]~~
=\lambda_{0}^{2}\zeta_{\tilde{\mathbf{H}}}(-2)~.
\end{equation}
As we shall see, the last two expressions  can be evaluated in  terms of the Hurwitz $\zeta$ function,  which is defined as the analytic continuation of the sum
\begin{equation}
\zeta(k,q)=\sum_{n=0}^{\infty}\frac{1}{(n+q)^{k}}~~.
\end{equation}


To illustrate the above procedure, the quantum systems we consider are the following: the first one is a generalised oscillator   \footnote{ We
	consider two different cases, namely when a constant driving force is present (which is 
	discussed in Appendix. \ref{timeindexamples} in detail) and the other one is when there is no driving force. The bi-invariant cost function 
		associated with a  generalised oscillator without a constant driving force was obtained \cite{YK}. Here we briefly review this procedure which we shall
		then generalise in the next section for time-dependent Hamiltonians.}
and the second one is an isotonic oscillator where, apart from a simple harmonic term,  an additional singular inverse square term is also present
in the total potential (this example is discussed in Appendix \ref{timeindexamples}). 
The CG generated by such Hamiltonians can be obtained by using the zeta-function regularization technique
and their implications will be discussed. All the systems considered in this paper are one-dimensional. 

\subsection{Complexity geometry generated by a generalised oscillator Hamiltonian}\label{regularHO}

We start with the Hamiltonian of  a generalised oscillator
\begin{equation}\label{generalosc}
\mathbf{H}_{GO}\big(\mathbf{X},\mathbf{P}\big)=\frac{A_{0}}{2}\mathbf{X}^{2}+\frac{B_{0}}{2}\mathbf{P}^{2}+\frac{C_{0}}{4}\big(\mathbf{X}\mathbf{P}+\mathbf{P}\mathbf{X}\big)~,
\end{equation}
where  $B_{0}\neq0$ and all the coefficients are constants, and $\mathbf{X}$ and $\mathbf{P}$ denote position and momentum operator respectively. 
The term with mixed position and momentum dependence can be removed by performing  the following unitary transformation on the above Hamiltonian
\begin{equation}\label{U1}
U_{1}(\mathbf{X})=\exp\bigg[i\frac{C_{0}}{4B_{0}}\mathbf{X}^{2}\bigg]~.
\end{equation}
It can be checked that, this unitary transformation is actually equivalent to a quantum canonical transformation of the operators 
$\mathbf{X},\mathbf{P}$ to a new pair of operators. After this  transformation, the Hamiltonian in \eqref{generalosc} reduces to
\begin{equation}\label{H1}
\mathbf{H}_{HO}=U_1(\mathbf{X})\mathbf{H}_{GO}(\mathbf{X},\mathbf{P})U_1^{\dagger}(\mathbf{X})=\frac{A_{1}}{2}\mathbf{X}^{2}+
\frac{B_{1}}{2}\mathbf{P}^{2}~,~~~~~\text{where}~~~A_{1}=A_{0}-\frac{C_{0}^{2}}{4B_{0}}~,~~\text{and}~~B_{1}=B_{0}~.
\end{equation}
This is just the Hamiltonian of a  HO with mass $m=1/B_{1}$ and frequency $\omega=\sqrt{A_{1}B_{1}}=\sqrt{A_{0}B_{0}-\frac{C_{0}^{2}}{4}}$.  Since the two Hamiltonians $\mathbf{H}_{GO}$ and $\mathbf{H}_{HO}$ are related by the unitary transformation $U_{1}$, due to bi-invariance,  the cost (\ref{p=1cost}) of the operators generated by both of them are equal, and hence, to find out the CG generated by the generalised oscillator in Eq. (\ref{generalosc}) we just need to evaluate the zeta function Eq. (\ref{zetaHdis}) for the HO Hamiltonian for different values of $k$.   In this case,  it  is given by
\begin{equation}
\zeta(k)_{\tilde{H}_{GO}}=\sum_{n=0}^{\infty}\Big(\omega \Big(n+\frac{1}{2}\Big)-a\Big)^{-k}=\omega^{-k}\zeta\Big(k,\frac{1}{2}-\frac{a}{\omega}\Big)~.
\end{equation}
In the second expression  the definition of the Hurwitz $\zeta$ function  has been used. When this is evaluated for $k=-1$  (with $\omega\neq0$), from the regularization criterion Eq. (\ref{regularization}), we have   the mean value $a$ to be  (see \cite{YK})
\begin{equation}\label{mean}
\zeta\Big(-1,\frac{1}{2}-\frac{a}{\omega}\Big)=0~, ~~~\text{i.e.}~~~a=\pm\frac{\omega}{2\sqrt{3}}~.
\end{equation}
With these values of the mean $a$, from Eq. (\ref{zetaHdis}) with $k=-2$, the bi-invariant cost is given by
\begin{equation}\label{GHOcost}
F^{2}_{r}=\lambda_{0}^{2}\omega^{2}\zeta\Big(-2,\frac{1}{2}-\frac{a}{\omega}\Big)=\frac{\sqrt{3}\lambda_{0}^{2}}{108}\omega^{2}
=\lambda_{1}^{2}\big(4A_{0}B_{0}-C_{0}^{2}\big)~,
\end{equation}
where $\lambda_{1}^{2}=\frac{\sqrt{3}\lambda_{0}^{2}}{432}$. 
Notice that since the cost can not be negative, from the two values of $a$, we need  to take only the positive one. As we shall see, this conclusion would change when we add a singular perturbation  to the HO Hamiltonian.

\subsection{Expression for the complexity }
Due to the bi-invariance of the CG, the calculation of NC is exceptionally simple for the quantum systems considered in this section since, for  these cases, the geodesic is 
generated by a constant generator \cite{Alexandrino } - here
the Hamiltonian. Thus, from the expression for the regularised norm derived in Eq. (\ref{regularisedcost}), we have the expression for the NC of a  given operator to be \cite{YANZK2}
\begin{equation}\label{BIcomp}
\mathcal{C}(\mathcal{O})=F(H_0)~~~~\text{where}~~~~\exp(-iH_0)=\mathcal{O}~.
\end{equation}

This completes our discussion of the CG of the operators generated by the time-independent quantum mechanical Hamiltonians,
and  we now turn to the cases where the Hamiltonian  explicitly depends on the time parameter.

\section{Complexity geometry of a harmonic oscillator with time-dependent frequency}\label{OTFCG}
From this section onward, we shall consider Hamiltonians which are explicitly time-dependent.  This kind of  Hamiltonian commonly appears when 
	the quantum system under consideration is influenced by some external perturbation (i.e., are not closed) so that the parameters in the Hamiltonian
	become functions of time. One example of such a system can be an atom undergoing radiative  transitions such that its interaction with the classical 
	radiation field contributes an explicitly time-dependent term in the Hamiltonian. For the purposes of the present paper, the situations that are relevant 
	are the systems that are driven far from equilibrium by some external perturbation,  a process commonly known as the quantum quenches.  When this 
	process occurs during a finite interval of time, the Hamiltonian governing the time evolution of the system becomes explicitly time-dependent. 
	See section \ref{quench} for an example of such quench protocol where we shall explicitly compute the cost derived later in this section and compare it with other
	information theoretic quantities, such as  change in the Shannon entropy. 
	
Our essential goal in this section is to calculate the regularised  cost function of the form
$F^{2}_{r}\big(\mathbf{H}(t),t\big)=\lambda_{0}^{2}\text{Tr}\big[\big(\mathbf{H}(t)-a(t)\mathbf{I}\big)^{2}\big]$.  Here we have introduced an additional argument 
$t$ in the expression for the cost to explicitly indicate that it is a function of time characterising the history of the evolution. 
We first obtain the diagonal and off-diagonal  matrix elements of a time-dependent Hamiltonian in a suitable  orthogonal basis, and use this to obtain the norm of the Hamiltonian which is the bi-invariant  cost in subsection  \ref{matrix} and \ref{mean-value}
respectively. This involves a regularisation procedure similar to the one discussed in the previous section, which we elaborate on in subsection \ref{com-geo}.
We finish this section with two illustrative analytical examples  in \ref{ex1} and \ref{ex2}  where the exact expression for the regularised cost can be written down in compact form.

\subsection{Matrix elements of the Hamiltonian}\label{matrix}
We start by considering one of the most commonly studied examples of quantum mechanical systems with explicit time dependence, namely
that  of a HO with time-dependent frequency (OTF for short from now on), whose Hamiltonian is given by 
\begin{equation}\label{OTF}
\mathbf{H}_{\Omega}(\mathbf{P},\mathbf{X},t)=\frac{1}{2}\Omega^{2}(t)\mathbf{X}^{2}+\frac{1}{2}\mathbf{P}^{2}~,
\end{equation}
where we have taken the mass of the oscillator to be unity. We also assume that the time-dependent frequency
$\Omega(t)$ is always a real function of time.  Our goal is to obtain the bi-invariant CG generated 
by the Hamiltonian of the OTF, i.e., we want  to calculate the trace of the displaced time-dependent Hamiltonian.  
 As before, we can add  a pure time-dependent term with the above Hamiltonian  from the beginning, and interpret it as the mean value used to  regularise the trace of this Hamiltonian, and since it only adds   a time-dependent pure phase factor in the total wavefunction,  the displaced Hamiltonian $\tilde{\mathbf{H}}_{\Omega}(t)$ and   the 
original one  $\mathbf{H}_{\Omega}(t)$ \footnote{From now on, in the notation for the Hamiltonian, we shall only show its explicit dependence on time and suppress other operators
	for convenience. } correspond to the same CG. 
The classical equation of motion derived from the above Hamiltonian is given by the usual second order equation
\begin{equation}\label{CEOMOTF}
	\frac{d^{2}x_c(t)}{dt^2}+\Omega^2(t)x_c(t)=0~.
\end{equation}

The quantum theory of an OTF has been studied  extensively in the literature.  It is well known that for such a system,  by using the   Lewis-Riesenfeld  method,   a  class of  exact time-dependent invariant operators  (denoted here as $\mathcal{I}(t)$)  can be found \cite{LR}. 
Solving the eigenvalue equation 
for such an invariant operator,  the  time-dependent   wavefunctions for an OTF can be obtained  by using the fact that these wavefunctions  are related to the eigenfunctions of the invariant operator by a time-dependent phase factor.  Therefore,  wavefunction of the $n$-th quantum state of the OTF ($\psi_{n} (x,t)$) are given in terms of the $n$th eigenfunctions ($\phi_{n}(x,t)$) of the corresponding invariant operator  as 
\begin{equation}\label{eigen}
\big|\psi_{n}(t)\big>=\exp\big[i\Theta_{n}(t)\big]\big|\phi_{n}(t)\big>~,
\end{equation}
where the state $\big|\phi_{n}(t)\big>$ satisfies the eigenvalue equation $\mathcal{I}(t)\big|\phi_{n}\big(t\big)\big>=\Lambda_{n}\big|\phi_{n}(t)\big>$, with $n=0,1,2\cdots$. By defining suitable creation and annihilation operators \footnote{We do not write down the explicit form for these operators which  can be found, e.g., in \cite{LR}, or can be obtained as a special case from the ones given in the next section in the context of the generalised time-dependent oscillator. See Eq. (\ref{aadag}) below. } to factorize   $\mathcal{I}$,  these eigenvalues can be found to be $\Lambda_{n}=\big(n+\frac{1}{2}\big)$. Furthermore, the expression for the 
time-dependent phase factor appearing in Eq. (\ref{eigen}) is given by \cite{LR,DR}
\begin{equation}
\Theta_{n}(t)=-\Big(n+\frac{1}{2}\Big)\int_{0}^{t}\frac{1}{\rho^{2}(t^{\prime})}dt^{\prime}~,
\end{equation}
where the auxiliary function $\rho(t)$ satisfies the following differential equation
\begin{equation}\label{auxiliary}
\ddot{\rho}(t)+\Omega^{2}(t)\rho(t)-\frac{1}{\rho(t)^{3}}=0~.
\end{equation}

For our purpose of calculating the trace of the displaced Hamiltonian, it is important to know the matrix elements of  $\mathbf{H}_{\Omega}$ with respect to $\big|\psi_{n}(t)\big>$. Since $\mathbf{H}_{\Omega}(t)$  does not contain any explicit time derivative, the diagonal matrix elements 
are the same as those with respect to $\big|\phi_{n}(t)\big>$. By using the same creation and annihilation operators as the once
 used to find out the eigenvalues of the invariant operator, these diagonal matrix elements, which we call $E_{nn}(t)$ for convenience, are given by 
\begin{equation}\label{diagonal1}
E_{nn}(t)=\big<\phi_{n}(t)\big|\mathbf{H}_{\Omega}(t)\big|\phi_{n}(t)\big>=\frac{1}{2}\Big(n+\frac{1}{2}\Big)f(t)~~~
\text{with}~~f(t)=\dot{\rho}^{2}(t)+\Omega^{2}(t)\rho(t)^2+\frac{1}{\rho(t)^{2}}~.
\end{equation}
 The non-diagonal elements $E_{mn}(t)$ (for   $m\neq n$) can  be computed by following a  similar procedure using the creation and annihilation 
 operator,  and are given by
\begin{equation}\label{nondiagonal1}
\begin{split}
E_{mn}(t)=\big<\phi_{m}(t)\big|\mathbf{H}_{\Omega}(t)\big|\phi_{n}(t)\big>=\frac{1}{4}\bigg\{\Big(g_{1}(t)-g_{2}(t)\Big)\sqrt{n\big(n-1\big)}\delta_{m+2,n}+\Big(g_{1}(t)+g_{2}(t)\Big)\sqrt{\big(n+2\big)\big(n+1\big)}\delta_{m,n+2}\bigg\}~~\\
\text{where}~~~g_{1}(t)=\dot{\rho}(t)^{2}-\rho(t)\ddot{\rho}(t)~,~ \text{and}~~g_{2}(t)=2i\frac{\dot{\rho}(t)}{\rho(t)}~.\hspace{2 cm}
\end{split}
\end{equation}
The derivations of the above expressions for the diagonal and non-diagonal elements can be found in \cite{LR},
 and we do not repeat it here for brevity. 

\textbf{The  function $f(t)$ and the instantaneous frequency.}
As we shall  see, the time-dependent  function $f(t)$ defined in Eq. \eqref{diagonal1} plays an important role in the derivation of 
the CG associated with the Hamiltonian $\mathbf{H}_{\Omega}(t)$. Therefore, before moving into  the calculation for the trace and CG, here we discuss a few properties of this function. Firstly, we note  the following  relation connecting the three time-dependent functions $f(t),g_{1}(t)$ and $g_{2}(t)$ defined above:
\begin{equation}\label{g12f}
g_{1}^{2}(t)-g_{2}^{2}(t)=f^{2}(t)-4\Omega^{2}(t)~.
\end{equation}
This can be easily verified by using the definition of these functions as given above, and the differential equation (\ref{auxiliary}) satisfied by the auxiliary function $\rho(t)$, and  will be very useful later on. Furthermore, it can also be seen that both the sides of this equality are strictly positive, since the coefficients of the two delta functions above in Eq.  (\ref{nondiagonal1}) are complex conjugates of each other, i.e., each side of \eqref{g12f} are just two different representations of the absolute value of a time-dependent 
complex function.

An useful integral expression of the function  $f(t)$ can be derived by multiplying Eq. (\ref{auxiliary}) by $\dot{\rho}(t)$, and   subsequently  integrating 
the resulting expression. This is given by
\begin{equation}\label{fint}
f(t)=\dot{\rho}^{2}(t)+\Omega^{2}(t)\rho(t)^2+\frac{1}{\rho(t)^{2}}=2\int \text{d}t~ \rho^2 \Omega\dot{\Omega} ~+c~,
\end{equation}
with $c$ being an integration constant. When the frequency of the oscillator is time-independent, this integral expression 
directly shows that $f(t)$ reduces to a constant. 

Next, using this integral expression it is possible to provide an interpretation of the $f$-function in terms of the instantaneous frequency
of the oscillator.  First, write the solution of the classical equation of motion  of the oscillator (Eq. (\ref{CEOMOTF}))   as 
\begin{equation}
	x_{c}(t)=\rho(t)\cos \theta(t)~.
\end{equation}
The time derivative of the function $\theta(t)$ is known as the instantaneous frequency of the OTF. 
Substituting this into the classical equation of motion we obtain a relation between the instantaneous frequency and the auxiliary function, namely, $\dot{\theta}(t)=1/\rho^{2}$ \cite{AD},\footnote{Here, we have set the angular momentum per unit mass of the oscillator to be unity.}  as well as a differential equation for the evolution of $\rho(t)$, which, as one can easily check is nothing but  the auxiliary equation given in Eq. (\ref{auxiliary}).\footnote{This is the reason of using the same notion $\rho$ both for this variable, as well as the auxiliary function appearing in the Lewis-Riesenfeld  invariant. } Furthermore, using these two relations, it can also be shown that the frequency $\Omega(t)$ can be written in terms of $\theta(t)$ as
\begin{equation}\label{SchD}
	\Omega^{2}(t)=\dot{\theta}(t)^{2}+\frac{1}{2}\{\theta(t),t\}~, ~~\text{where}~~\{\theta(t),t\}=\frac{\dddot{\theta}(t)}{\dot{\theta}(t)}-\frac{3}{2}\Big(\frac{\ddot{\theta}(t)}{\dot{\theta}(t)}\Big)^{2}~.
\end{equation} 
The quantity $\{\theta(t),t\} $ is known as the Schwarzian derivative,  and vanishes only when $\theta(t)$ is a linear fractional function of $t$. The presence of the second term in the above expression indicates that the time variation of the frequency is different from the instantaneous frequency ($\dot{\theta}$), and hence the 
instantaneous frequency  can not follow the time-dependent frequency ($\Omega(t)$) of the oscillator exactly.

Now substituting the expression for $\Omega(t)$ into the integral representation of $f(t)$ in Eq. \eqref{fint} we obtain
\begin{equation}
	f(t)=2\int \ddot{\theta}(t)~\text{d}t + \frac{1}{2}\int ~\rho(t)^{2}\frac{d}{dt}\Big(\{\theta(t),t\}\Big) ~\text{d}t+c~.
\end{equation}
First consider the case when   $\theta$ is a linear fractional transformation, so that $\{\theta(t),t\}=0$. When this condition is satisfied, the frequency of the oscillator  matches exactly with the instantaneous frequency. In that case, it is easy to see that the function 
$f(t)$, given by
\begin{equation}\label{fSD=0}
	f(t)=\frac{2}{\rho(t)^{2}}+c~=2\Omega(t)+c~,
\end{equation}
is directly  related to the time-dependent frequency. To understand the significance of this result, we notice that, when the Schwarzian derivative
vanishes, it is possible to obtain an easy expression for the auxiliary function. Using  the relation $\dot{\theta}(t)=1/\rho^{2}(t)$ in the expression for the Schwarzian derivative
in Eq. \eqref{SchD}, 
we see that condition that it vanishes is equivalent to  $\ddot{\rho}(t)=0$. Thus in this case $\dot{\rho}(t)=l_{1}$ is a constant and $\rho(t)=l_{1}t+l_{2}$ is just a linear function of time. The validity of Eq. (\ref{fSD=0}) can now also be directly checked using the definition of $f(t)$ given in Eq. (\ref{diagonal1}) as well. Therefore, when the time-dependent frequency of the oscillator is of the form $\Omega(t)=1/(l_{1}t+l_{2})^{2}$, the auxiliary function is a linear function of time, and the instantaneous frequency perfectly follows the original time evolution. We shall come back to this important example later on after deriving the CG.

\subsection{Calculation of mean value of the Hamiltonian} \label{mean-value}
We now go back to the  computation of  the mean value $a(t)$, which is obtained  by imposing the criterion Eq. (\ref{regularization}) on the Hamiltonian in Eq. (\ref{OTF}) i.e. $\text{Tr}_{r}\big[\mathbf{H}_{\Omega}(t)-a(t)\mathbf{I}\big]=0$. Since the Hamiltonian is an explicit function of time,  it is expected that the mean would generally depend on time, i.e., $a=a(t)$.
To calculate the regularised trace, we need the diagonal matrix elements of  the displaced Hamiltonian $\tilde{\mathbf{H}}_{\Omega}(t)$ 
with respect to the states $\big|\psi_{n}(t)\big>$. Since $\tilde{\mathbf{H}}_{\Omega}(t)$ does not contain any explicit time derivative, these are 
given by (from Eq. (\ref{diagonal1}))
\begin{equation}
\tilde{E}_{nn}(t)=\big<\psi_{n}(t)\big|\tilde{\mathbf{H}}_{\Omega}(t)\big|\psi_{n}(t)\big>=E_{nn}(t)-a(t)=\frac{1}{2}\Big(n+\frac{1}{2}\Big)f(t)-a(t)~.
\end{equation}
 By using a similar procedure as that of the HO, we can define the zeta function corresponding to that of $\tilde{\mathbf{H}}_{\Omega}(t)$ as the analytical continuation of the following sum
\begin{equation}
\zeta_{\tilde{\mathbf{H}}_{\Omega}}(k)=\sum_{n=0}^{\infty}\frac{1}{\big(E_{nn}(t)-a(t)\big)^{k}}~~=\bigg(\frac{f(t)}{2}\bigg)^{-k}\sum_{n=0}^{\infty}\bigg( \Big(n+\frac{1}{2}\Big)-\frac{2a(t)}{f(t)}\bigg)^{-k}~=\bigg(\frac{f(t)}{2}\bigg)^{-k}\zeta\bigg(k,\frac{1}{2}-\frac{2a(t)}{f(t)}\bigg)~.
\end{equation}
We now choose $a(t)$ in such a way that the quantity $a(t)/f(t)$ is a constant which we call $\gamma$. As before,  the regularisation criterion fixes this constant to be
\begin{equation}
\zeta\Big(-1,\frac{1}{2}-2\gamma\Big)=0~~~\rightarrow~~~\gamma=\frac{a(t)}{f(t)}=\pm\frac{1}{4\sqrt{3}}~.
\end{equation}

\subsection{The complexity geometry}\label{com-geo}
We now turn to the calculation of the norm of  this Hamiltonian, which is later used to obtain the CG of the operator generated by it. For this, we need to evaluate the quantity
\begin{equation}
F^{2}_{r}\big(\mathbf{H}_{\Omega}(t)\big)=\lambda_{0}^{2}\text{Tr}_{r}\Big[\Big(\mathbf{H}_{\Omega}(t)-a(t)\mathbf{I}\Big)^{2}\Big]~.
\end{equation}
Contrary to the  previous cases of the time-independent Hamiltonians,  since in this case there are nonzero non-diagonal matrix elements of $\mathbf{H}_{\Omega}(t)$,  it is profitable to write the  above quantity explicitly in terms of the  basis of the states $\big|\psi_{n}(t)\big>$. This expansion is given by
\begin{equation}\label{NormH5}
\begin{split}
F^{2}_{r}\big(\mathbf{H}_{\Omega}(t)\big)=\lambda_{0}^{2}\sum_{n,m}\big<\psi_{n}(t)\big|\big(\mathbf{H}_{\Omega}(t)-a(t)\mathbf{I}\big)\big|\psi_{m}(t)\big>\big<\psi_{m}(t)\big|\big(\mathbf{H}_{\Omega}(t)-a(t)\mathbf{I}\big)\big|\psi_{n}(t)\big>~\\
=\lambda_{0}^{2}\sum_{n}\Big\{\big<\psi_{n}(t)\big|\big(\mathbf{H}_{\Omega}(t)-a(t)\mathbf{I}\big)\big|\psi_{n}(t)\big>\Big\}^{2}+\lambda_{0}^{2}\sum_{n,m\neq n}\big<\psi_{n}(t)\big|\mathbf{H}_{\Omega}(t)\big|\psi_{m}(t)\big>\big<\psi_{m}(t)\big|\mathbf{H}_{\Omega}(t)\big|\psi_{n}(t)\big>~.
\end{split}
\end{equation} 
The first term above, which we  call $T_{1}$ for convenience,  is evaluated following  the method given for the time-independent Hamiltonians 
in the previous section, and utilising the diagonal matrix elements given in Eq. (\ref{diagonal1}) we can  express it as 
\begin{equation}\label{term1}
T_{1}=\lambda_{0}^{2}\sum_{n}\Big(E_{nn}-a(t)\Big)^{2}=\lambda_{0}^{2}\bigg(\frac{f(t)}{2}\bigg)^{2}\zeta\bigg(-2,\frac{1}{2}+\frac{1}{2\sqrt{3}}\bigg)~=\lambda_{1}^{2}f^{2}(t)~,
\end{equation}
where in the last expression we have defined the constant $\lambda_{1}^{2}=\frac{\sqrt{3}}{432}\lambda_{0}^{2}~$.  Notice that, in writing the 
final form of $T_1$ we have chosen the negative value of $\gamma$. However,  since the expression for the  norm of the Hamiltonian in
Eq. \eqref{NormH5} actually given by the addition of two terms, and as we shall see in a moment,  the second term is positive; therefore, even if we choose the positive value of $\gamma$, the resulting cost could come out to be positive. We can resolve this ambiguity by imposing
the natural condition that the final expression of the norm of $\mathbf{H}_{\Omega}(t)$ (which we derive below) should reduce to that of HO in the limit of constant frequency. As we shall see below, in the limit of constant frequency, the second term in Eq. (\ref{NormH5}) vanishes, so that to make this expression positive in this limit, and hence, to match with that of the HO result,  we need to choose $\gamma$ to be negative. 

To evaluate the second term $T_{2}$ we use the nondiagonal matrix elements of $\mathbf{H}_{\Omega}(t)$ given in Eq. (\ref{nondiagonal1}) and proceed as follows 
\begin{equation}\label{T2}
\begin{split}
T_{2}=\frac{\lambda_{0}^{2}}{16}\Big(g_{1}^{2}(t)-g_{2}^{2}(t)\Big)\sum_{n,m\neq n}\bigg\{\sqrt{n\big(n-1\big)}\sqrt{(m+1)\big(m+2\big)}\delta_{m,n-2}\delta_{n,m+2}\\
+\sqrt{\big(n+1\big)\big(n+2\big)}\sqrt{m\big(m-1\big)}\delta_{m,n+2}\delta_{n,m-2}\bigg\}~.
\end{split}
\end{equation}
Notice that the relative phase difference between different terms, of the form    $\exp[\Theta_{mn}]=\exp[\Theta_{m}-\Theta_{n}]$, cancels
with each other in the above expression. After simplification, we finally obtain
\begin{equation}\label{T2fi}
	\begin{split}
		T_{2}
		=\frac{\lambda_{0}^2}{8}\Big(g_{1}^{2}(t)-g_{2}^{2}(t)\Big)\sum_{n=0 }^{\infty}\Big(n^{2}+n+1\Big)~~=\frac{\lambda_{0}^2}{8}\Big(f^{2}(t)-4\Omega^{2}(t)\Big)\sum_{n=0 }^{\infty}\Big(\big(n+\frac{1}{2}\big)^2+\frac{3}{4}\Big)~.
	\end{split}
\end{equation}
The series in this expression is an example of the generalised Epstein-Hurwitz series,  defined as \cite{EE}
\begin{equation}
	E_{1}(s,\alpha,\beta)=\sum_{n=0 }^{\infty}\Big(\big(n+\alpha\big)^2+\beta^2\Big)^{-s}~.
\end{equation}
What we need here to evaluate the cost is the numerical value of this series for $\alpha=1/2,\beta^2=3/4,s=-1$. 
Such series and its derivatives are well studied in the literature; see, e.g., \cite{EE} for an extensive review. The asymptotic expansions of this series both for large and small values of the parameter $\beta$ are available which provide good numerical approximations of the series. 

Finally substituting the expressions in Eq. (\ref{term1}) and Eq. (\ref{T2fi}) into Eq. (\ref{NormH5}) we obtain the following compact expression for  cost function 
\begin{equation}\label{NormH5final}
	F^{2}_{r}\big(\mathbf{H}_{\Omega}(t)\big)=\lambda_{1}^{2}f^{2}(t)+\lambda_{2}^{2}\Big(f^{2}(t)-4\Omega^{2}(t)\Big)~=~\lambda_{12}^{2}f^{2}(t)-4\lambda_{2}^{2}\Omega^{2}(t)~,
\end{equation}
where $\lambda_{1}^2$ and $\lambda_{2}^2$ are two constants that incorporate the numerical factors from the first and second terms, respectively, and 
we have also renamed $\lambda_{12}^{2}=\lambda_{1}^{2}+\lambda_{2}^{2}$ in the second expression. This constitutes one of the main results 
of the present paper. 
Clearly, for a normal oscillator, for which the frequency is always a real function of time, the role of the last term   in the  second expression above (which is a part of the contribution of  the non-diagonal matrix elements of the Hamiltonian towards the norm)  is to reduce the total cost. An alternative expression for the cost function can be obtained  in terms of the expectation value of the Hamiltonian in the state $\big|\phi_{0}(t)\big>$  (and the frequency)   by noting that the function $f(t)$ is related to it by   Eq. (\ref{diagonal1})
so that we have, 
\begin{equation}
	F^{2}_{r}\big(\mathbf{H}_{\Omega}(t)\big)=16\lambda_{12}^{2}\Big(\big<\phi_{0}(t)\big|\mathbf{H}_{\Omega}(t)\big|\phi_{0}(t)\big>\Big)^{2}-4\lambda_{2}^{2}\Omega^{2}(t)~.
\end{equation}

After deriving this expression for the cost we need to show that in an appropriate limit this reduces to the expression for the HO obtained in \cite{YK} and discussed in section \ref{regularHO}.  To check this,  we notice that in the limit $\Omega(t)=\omega=\text{constant}$, it is possible to directly integrate Eq. (\ref{auxiliary}) to obtain a first integral of motion (this can be directly seen from the general formula derived in Eq. (\ref{fint}) by putting $\dot{\Omega}=0$). With  the  integration constant written as $2\omega$ (with $\omega>0$) we have \cite{LR}
\begin{equation}\label{1stintegral}
f(t)=\bigg(\dot{\rho}^{2}(t)+\Omega^{2}(t)\rho(t)^{2}+\frac{1}{\rho(t)^{2}}\bigg)_{\Omega(t)=\omega}=2\omega~,
\end{equation}
so that the second term in the  (first expression of the) cost in Eq. (\ref{NormH5final}) vanishes, and we recover the result obtained in \cite{YK}, i.e., 
 that the cost is proportional to the square of the frequency of a HO.
It is important to note that the constant $2\omega$ is not the most general choice one can make. In  fact, it is possible to write the most general first integral of Eq. (\ref{auxiliary}) at the limit of constant  $\Omega$ to be equal to  $f(t)=2\omega \cosh \delta$.  However, as it has been shown in \cite{LR}, 
 if one assumes that in the limit  $\Omega(t \rightarrow \pm \infty )=\omega$,  the auxiliary function reduces to  a constant, i.e., 
  $\rho (t \rightarrow \pm \infty)=\omega^{-1/2}$, it necessarily implies $\delta=0$, and we recover $f(t)=2\omega$ in the limit 
  of constant frequency.

\textbf{A bound on the cost:}
We now derive  an interesting bound on the total cost of the time evolution  by using the Cauchy-Schwarz inequality for  square-integrable functions. Substituting the expression for  the cost function from Eq. (\ref{NormH5final})  in the expression for the cost functional, with the endpoint being  an arbitrary point  on the time evolution curve (denoted as $T$), we can  proceed as follows, 
\begin{multline}
	\big(\mathcal{C}_{\Omega}(T)\big)^{2}=\bigg[\int_{0}^{T}\sqrt{\lambda_{12}^{2}f^{2}(t)-4\lambda_{2}^{2}\Omega^{2}(t)} ~\text{d}t\bigg]^{2}\\
	~\leq~T\bigg[\int_{0}^{T} \bigg(\lambda_{12}^{2}f^{2}(t)-4\lambda_{2}^{2}\Omega^{2}(t) \bigg) ~\text{d}t\bigg]~\leq~ \lambda_{12}^{2}T\int_{0}^{T}~f^{2}(t)~\text{d}t =
	\lambda_{12}^{2} T^{2}\overline{f^{2}(t)}~,
\end{multline}
with $\overline{f^2}$ denoting the  time average of the function $f(t)$ up to time $T$. Combining the final  and the initial expressions above, we have, for an arbitrary point on the unitary 
evolution curve generated by a  OTF, the following upper  bound on the total cost 
\begin{equation}\label{bound}
	\mathcal{C}_{\Omega}(T)~\leq~\lambda_{12}T \sqrt{\overline{f^{2}(t)}}~=~\lambda_{12}T\sqrt{\overline{\Big(\big<\phi_{0}(t)\big|\mathbf{H}_{\Omega}(t)\big|\phi_{0}(t)\big>\Big)^{2}}}~.
\end{equation}
Therefore, the upper bound of the total cost is proportional to the square root  of time average of the  square of the expectation value of the Hamiltonian in the lowest 
quantum state of the oscillator. 
Furthermore, we can also formally  invert the above expression to obtain a lower bound on the time required to reach a  given value of total  cost:
\begin{equation}
	T~\geq~\frac{\mathcal{C}_{\Omega}(T)}{\lambda_{12} \sqrt{\overline{f^{2}(t)}}}~=~\frac{	\mathcal{C}_{\Omega}(T)}{\lambda_{12} \sqrt{\overline{\Big(\big<\phi_{0}(t)\big|\mathbf{H}_{\Omega}(t)\big|\phi_{0}(t)\big>\Big)^{2}}}}~.
\end{equation}
Inequality such as this often appears in the context of the quantum speed limit \cite{Deffner:2017cxz},  which is usually 
defined as  the minimum amount of time required to prepare a quantum state with unit fidelity. 
See, e.g. \cite{Bukov},  where such a geometric bound for the quantum speed limit in quantum many-body systems has been conjectured.



To illustrate the results presented so far in this section, we now consider some OTF models that are commonly studied  in the literature.  
Later in section \ref{quench}, we discuss one further example of a time-dependent protocol, namely a quantum quench 
scenario, which is one of the most popular protocols for studying non-equilibrium dynamics in quantum systems, and find out
the behaviour of different physical quantities in adiabatic as well as fast quench regimes. 

\subsection{Example-1 : Instantaneous frequency with non-vanishing Schwarzian derivative  }\label{ex1}

As the  first example consider  the following time-dependent  frequency profile
\begin{equation}
\Omega(t)=\frac{1}{b_{1}+b_{2}t}~,
\end{equation}
where $b_{1}$ and $b_{2}$ are two positive real constants. For this frequency profile we need a solution of the auxiliary Eq. (\ref{auxiliary}) 
for two different ranges  for the constant $b_{2}$,  namely $ 0<b_{2}<2$ and $b_{2}=2$ \cite{SAHM}.

\textbf{Case-1 : $ 0<b_{2}<2$ .}
In this case, the solution for Eq. (\ref{auxiliary}) is given by
\begin{equation}
\rho(t)=c_{1}\sqrt{b_{1}+b_{2}t}~~,~~ \text{with}~~~c_{1}=\Big(1-\frac{b_{2}^{2}}{4}\Big)^{-1/4}~.
\end{equation}
Clearly, this solution is not valid for the value of the constant $b_{2}=2$ for which $c_{1}=0$, and therefore we need to consider this case separately (as we describe below).  
Using this solution in  Eq. (\ref{NormH5final}), it is easy to see that the norm of the Hamiltonian is 
\begin{equation}\label{ex11}
F^{2}_{r}\big(\mathbf{H}_{\Omega}(t)\big)=\frac{16\lambda_{1}^{2}+4(b_{2}\lambda_{2})^{2}}{\big(4-b_{2}^{2}\big)\big(b_{1}+b_{2}t\big)^{2}}~
=\big(16\lambda_{1}^{2}+4(b_{2}\lambda_{2})^{2}\big)\frac{\Omega^{2}(t)}{\big(4-b_{2}^{2}\big)}
=\lambda_{3}^{2}\Omega^{2}(t)~,
\end{equation}
where we have defined the new constant $\lambda_{3}$ whose form  can be obtained by comparing the last two expressions above.
Therefore, the norm of the Hamiltonian is proportional to the square of the time-dependent frequency - an observation that will be important later on. Moreover, in the limit of infinite time, the frequency vanishes, making the Hamiltonian that of a free particle, so that the norm also goes to zero in that limit. 

For this particular example, the limit of a  HO is easy to check as well; this  is just the limit $b_{2}\rightarrow 0$. In this limit $\Omega \rightarrow 1/b_{1}$ and  $\rho(t) \rightarrow \sqrt{b_{1}}$,   so that $f(t)=2/b_{1}$  and  thus  the condition in Eq. (\ref{1stintegral}) is satisfied,  and the expression for the cost
 reduces to that of the HO with frequency $ \omega=1/b_{1}$. Clearly, the actual values of the constants $\lambda_{1}^{2}$ and $\lambda_{2}^{2}$ are not important in this case since they only affect the overall magnitude of the norm, not its time dependence,  and the bi-invariant construction only determines it upto this overall constant multiplier.


The exact expression for the total cost can also be obtained by substituting the above cost function in Eq. (\ref{complexitydef}) with the endpoint 
taken as a general point (denoted as $t_f$) of time evolution 
\begin{equation}
	\mathcal{C}_{\Omega}(t_f)=\int_{0}^{t_f}~F_{r}\big(\mathbf{H}_{\Omega}(t)\big)~\text{d}t
	= \frac{\lambda_{3}}{b_2}\ln \big[1+\frac{b_2}{b_1} t_f\big]~.
\end{equation}
Therefore, for large values of time, the total cost evaluated along the time evolution curve grows logarithmically. Here  the total cost is actually proportional to the average frequency of the oscillator. It can also be checked, using the definition given in sec - \ref{matrix},
for this example, the instantaneous frequency ($\dot{\theta}$) is different from the time-dependent frequency of the oscillator, 
i.e. the Schwarzian derivative of the phase $\theta(t)$ is non-zero.

\textbf{Case-2 : $b_2=2$.}
In this case a solution for the Eq. (\ref{auxiliary}) can be written as  \cite{SAHM}
\begin{equation}
\rho(t)=\sqrt{b_{1}+2t}\bigg\{1+\Big(\ln \sqrt{b_{1}+2t}\Big)^{2}\bigg\}^{1/2}~,
\end{equation}
so that the expression for the  cost is  given by
\begin{equation}
F^{2}_{r}\big(\mathbf{H}_{\Omega}(t)\big)=\frac{1}{\big(b_{1}+2t\big)^{2}}\Big[\mathcal{G}(t) \big(\lambda_{1}^{2}+\lambda_{2}^{2}\big)-\lambda_{2}^{2}\Big]~,
~~~\text{where}~~\mathcal{G}(t)=\frac{1}{4}\bigg(6+2\ln\big[b_{1}+2t\big]+\Big(\ln\big[b_{1}+2t\big]\Big)^{2}\bigg)^{2}~.
\end{equation}
Once again, we see that this expression is proportional to $\Omega^{2}(t)$; however, unlike the previous case, the 
time dependence is not solely given by the frequency. An analytical  expression for the total cost can be calculated as before, though this expression
is quite lengthy, and  we do not reproduce it here for brevity.

\subsection{Example-2 : Instantaneous frequency with vanishing  Schwarzian derivative  }\label{ex2}

As the second example, we come back to the special case discussed towards the end of sec-  \ref{matrix}, i.e., when the time-dependent  frequency is taken to be 
of the following form
\begin{equation}
	\Omega(t)=\frac{1}{(l_{1}t+l_{2})^{2}}~,
\end{equation}
where $l_1$ and $l_2$ are two positive real constants.  In this case, the Schwarzian derivative in Eq. \eqref{SchD} vanishes, and the expression for the  function $f(t)$ has been already obtained in Eq. (\ref{fSD=0}). Substituting this in the expression for the norm we have 
\begin{equation}
	F^{2}_{r}\big(\mathbf{H}_{\Omega}(t)\big)=\lambda_{1}^{2}(2\Omega(t)+c)^{2}+\lambda_{2}^{2}\Big((2\Omega(t)+c)^{2}-4\Omega^{2}(t)\Big)~
	=\lambda_{12}^2  (2\Omega(t)+c)^{2}- 4\lambda_{2}^2 \Omega^2(t)~.
\end{equation}
This expression has been written formally in terms of  $\Omega(t)$, however, unlike the one in Eq. \eqref{ex11},
the expression for the norm is not entirely proportional to the time-dependent 
frequency.\footnote{The norm is proportional to the frequency $\Omega(t)$ only in the special cases when the integration constant $c$ in Eq. \eqref{fSD=0} vanishes.}
In this example it is  also possible  to compute the total cost up to a time $t_f$ of time evolution by integrating the bi-invariant cost,
however, the analytical formula for it is quite lengthy, and therefore, we omit  it here.

\section{Complexity geometry of  a generalised time-dependent oscillator } \label{OTMFCG}

\subsection{Computation of the norm}\label{HOMFnorm}
In the previous section we have studied the CG generated by the Hamiltonian of an OTF and discussed a few examples where the corresponding bi-invariant 
cost can be evaluated. 
In this section we shall construct the CG generated by the Hamiltonian of the generalised time-dependent oscillator of the following form
\begin{equation}\label{generalosctime}
\mathbf{H}_{GO}\big(\mathbf{X},\mathbf{P},t\big)=\frac{A_{0}(t)}{2}\mathbf{X}^{2}+\frac{B_{0}(t)}{2}\mathbf{P}^{2}+\frac{C_{0}(t)}{4}\big(\mathbf{X}\mathbf{P}
+\mathbf{P}\mathbf{X}\big)~.
\end{equation}
As we show later in this section, the cost associated with this Hamiltonian has a close connection with that of  an OTF considered in the previous section. 
This Hamiltonian in \eqref{generalosctime} is of the same form as that of (\ref{generalosc}) considered previously,  but now   all the three constants $A_{0}(t),B_{0}(t),C_{0}(t)$ depend on time. We assume the following time dependence for  $B_{0}(t)$ and  $C_{0}(t)$: $B_{0}(t)=c_1 g(t) ,C_{0}(t)=c_2 g(t)$, where $ c_1 $ and $ c_1 $ are two constants and $ g(t)$ is   an arbitrary function of time. These forms for the coefficients have been chosen so that  
the ratio of these two coefficients is  a constant. The reason for this particular restriction on these coefficients will be explained shortly.

Let us now consider the following unitary operator 
\begin{equation}\label{U1t}
U_{1}=\exp\bigg[i\frac{C_{0}(t)}{4B_{0}(t)}\mathbf{X}^{2}\bigg]~=\exp\bigg[i\frac{c_1}{4c_2}\mathbf{X}^{2}\bigg]~,
\end{equation}
which transforms  the above Hamiltonian to a new one of the  form 
\begin{equation}\label{H1t}
\mathbf{H}_{m_1\Omega_1}\big(\mathbf{X},\mathbf{P},t\big)=\frac{A_{1}(t)}{2}\mathbf{X}^{2}+\frac{B_{1}(t)}{2}\mathbf{P}^{2}~,
~~~~~\text{where}~~~A_{1}(t)=A_{0}(t)-\frac{C_{0}^{2}(t)}{4B_{0}(t)}~~\text{and}~~B_{1}(t)=B_{0}(t)~.
\end{equation}
 We see that the transformed Hamiltonian is the same as that of Eq. (\ref{H1}) and the unitary transformation is the same as that of Eq. (\ref{U1}),
and the resulting Hamiltonian is that of the HO whose    mass ($m_{1}(t)=1/B_{1}(t)$) and frequency ($\Omega_1(t)=\sqrt{A_{1}(t)B_{1}(t)}$) both depend on time. The reason for choosing those particular forms of the functions  $B_{0}(t)$ and  $C_{0}(t)$ is that, for this choice,  the above unitary transformation is time-independent, and as a result, the 
transformed  Hamiltonian is of the form of a HO with time-dependent mass and frequency (OTMF).
For a general time-dependent transformation the final form of the Hamiltonian would have been different. 
Furthermore, it is important to note that since the two Hamiltonians in Eqs. (\ref{H1t}) and (\ref{generalosctime}) are related by the time-independent unitary transformation $U_{1}$,  due to unitary invariance of the CG in Eq. (\ref{p=1cost}), the norms of both the Hamiltonians are equivalent.  For a unitary transformation, which is 
a general function of time, the bi-invariant cost would no longer be invariant.  

The quantum theory of an  oscillator whose  mass and frequency are functions of time is well-studied in the literature. Once again, we can  use the Lewis-Riesenfeld  method   of finding  exact time-dependent invariant operators  ($\mathcal{I}(t)$) from an explicitly time-dependent Hamiltonian. Using the operator formalism, the eigenvalue problem  of such an operator can be solved and, in turn, can be used to find the solutions of the time-dependent Schrodinger equation for  the Hamiltonian in Eq. (\ref{H1t}). See,  e.g.,  \cite{IAP,CK} for further  details of this procedure - we shall use only the relevant results here.

We introduce the following creation and annihilation operators, written in terms of an auxiliary function ($\rho(t)$),
\begin{equation}\label{aadag}
\mathcal{A}=\frac{1}{\sqrt{2}}\bigg[\bigg(\frac{1}{\rho(t)}-i\frac{\dot{\rho}(t)}{B_{1}(t)}\bigg)\mathbf{X}+i\rho(t)\mathbf{P}\bigg]~,
~\mathcal{A}^\dagger=\frac{1}{\sqrt{2}}\bigg[\bigg(\frac{1}{\rho(t)}+i\frac{\dot{\rho}(t)}{B_{1}(t)}\bigg)\mathbf{X}-i\rho(t)\mathbf{P}\bigg]~,
\end{equation}
where the auxiliary function now satisfies the  following differential equation \cite{IAP,CK}
\begin{equation}\label{auxiliary2}
\ddot{\rho}(t)-\frac{\dot{B_{1}}(t)}{B_{1}(t)}\dot{\rho}(t)+A_{1}(t)B_{1}(t)\rho(t)-\frac{B_{1}^2(t)}{\rho(t)^{3}}=0~.
\end{equation}
These operators satisfy the usual commutation relation $[\mathcal{A}, \mathcal{A}^\dagger]=1$. 
Note that we have denoted the auxiliary function by $\rho(t)$ in this case also for convenience,   but it  should not be confused with that of the 
one used previously in the context of OTF, since that satisfies the differential equation in (\ref{auxiliary}), which is  different from the one written above. 
Inverting the relations in Eq. (\ref{aadag}) the position and 
momentum operators can be written in terms of the creation and annihilation operators, and these  can be subsequently substituted in the Hamiltonian 
in Eq. (\ref{H1t}) to get the following expression for it 
\begin{equation}\label{H1tra}
\begin{split}
\mathbf{H}_{m_1\Omega_1}\big(\mathbf{X},\mathbf{P},t\big)=\frac{1}{4}\Big[f_1(t)\mathcal{A}^2+f_2(t)(\mathcal{A}^\dagger)^2+f_3(t)(\mathcal{A}\mathcal{A}^\dagger+\mathcal{A}^\dagger \mathcal{A})\Big]~,
\end{split}
\end{equation}
where we have defined three time-dependent functions through the relations
\begin{equation}\label{fs}
	\begin{split}
		f_1(t)=A_{1}(t)\rho^{2}-\biggl\{B_{1}(t)-\frac{(\rho\dot{\rho})^2}{B_{1}(t)}+2i\rho\dot{\rho}\biggr\}\frac{1}{\rho^{2}}~,\\
		~f_2(t)=A_{1}(t)\rho^{2}-\biggl\{B_{1}(t)-\frac{(\rho\dot{\rho})^2}{B_{1}}-2i\rho\dot{\rho}\biggr\}\frac{1}{\rho^{2}}~,~~\text{and}~~\\
		f_3(t)=\frac{\dot{\rho}^{2}}{B_{1}(t)}+A_{1}(t)\rho^2+\frac{B_{1}(t)}{\rho^{2}}~.
	\end{split}
\end{equation}

Since all the time-dependent coefficients of the Hamiltonian are real, and the function $\rho(t)$ is real as well (we need to  choose the real solution of the auxiliary equation),  
the functions $f_{1}(t)$ and $f_{2}(t)$ are actually  complex conjugate to each other. This indicates that their product $f_{1}(t)f_{2}(t)$ is a real quantity -  a fact that will be 
useful later on.  With this form of the Hamiltonian, and using the standard properties of the  creation and annihilation operators when acted on an eigenstate $\big|\phi_{n}(t)\big>$ of the invariant operator, it is now easy to calculate the matrix elements of $\mathbf{H}_{m_1\Omega_1}$. The diagonal and nondiagonal matrix elements  are respectively given by  \cite{CK}
\begin{equation}
\begin{split}
\big<\phi_{n}(t)\big|\mathbf{H}_{m_1\Omega_1}(t)\big|\phi_{n}(t)\big>=\frac{1}{2}\Big(n+\frac{1}{2}\Big)f_{3}(t)~,~~~~\text{and}~~\hspace{2.4 cm}\\
\big<\phi_{m}(t)\big|\mathbf{H}_{m_1\Omega_1}(t)\big|\phi_{n}(t)\big>=\frac{1}{4}\bigg(f_{1}(t)\sqrt{n\big(n-1\big)}\delta_{m+2,n}+f_{2}(t)\sqrt{\big(n+2\big)\big(n+1\big)}\delta_{m,n+2}\bigg)
~, ~~~\text{for}~m\neq n~.
\end{split}
\end{equation}
For later purposes we record an important  relation between the time-dependent functions 
$f_1,f_2$ and $f_{3}$ defined above:
\begin{equation}\label{f123}
f_{1}(t)f_{2}(t)=f_{3}^{2}(t)-4A_{1}(t)B_{1}(t)~.
\end{equation}
This  can be easily checked by using the differential equation in Eq. (\ref{auxiliary2}) satisfied by the  
auxiliary  function.
Notice that, all the relations derived so far in this section reduced to that of those obtained in the previous section for an OTF,
when we put $B_{1}=1$  and appropriately adjust the notations.

Following the general procedure described  in the previous section we now proceed to the calculation of the norm
\begin{equation}
F^{2}_{r}\big(\mathbf{H}_{m_1\Omega_1}(t)\big)=\lambda_{0}^{2}\text{Tr}_{r}\Big[\Big(\mathbf{H}_{m_1\Omega_1}(t)-a(t)\mathbf{I}\Big)^{2}\Big]~,
\end{equation}
and arrive at a relation analogous to the second expression of Eq. (\ref{NormH5}).  Calling the first term as $T_{1}$ as before, we obtain its value to be
\begin{equation}\label{term11}
T_{1}=\lambda_{0}^{2}\bigg(\frac{f_{3}(t)}{2}\bigg)^{2}\zeta\Big(-2,\frac{1}{2}+\frac{1}{2\sqrt{3}}\Big)~=\lambda_{1}^{2}f_{3}^{2}(t)~.
\end{equation}
Note that, we have used the regularisation criterion $\text{Tr}_{r}\big[\mathbf{H}_{m_1\Omega_1}(t)-a(t)\mathbf{I}\big]=0$ to fix the the function $a(t)$ 
such that $\gamma=\frac{a(t)}{f_3(t)}$ is a constant, and equal to $\pm\frac{1}{4\sqrt{3}}$, and have subsequently chosen negative value of $\gamma$. 
Similarly, the final expression for the second term, after some algebraic simplifications are made and the relation in Eq. (\ref{f123}) is used,  can be written as 
\begin{equation}
\begin{split}
T_{2}
=\frac{\lambda_{0}^2}{8}f_{1}(t)f_{2}(t)\sum_{n=0 }^{\infty}\Big(n^{2}+n+1\Big)~~
=\frac{\lambda_{0}^2}{8}\Big(f_{3}^{2}(t)-4A_{1}(t)B_{1}(t)\Big)\sum_{n=0 }^{\infty}\bigg(\Big(n+\frac{1}{2}\Big)^2+\frac{3}{4}\bigg)~.
\end{split}
\end{equation}
The infinite summation above  is once again  an example of the  generalised Epstein-Hurwitz series, and we call its numerical value (including the factor of $\lambda_{0}^{2}$) to be $\lambda_{2}^{2}$, so that the expression for the norm is given by 
\begin{equation}\label{NormH1final}
F^{2}_{r}\big(\mathbf{H}_{m_1\Omega_1}(t)\big)=\lambda_{1}^{2}f^{2}_{3}(t)+\lambda_{2}^{2}\Big(f^{2}_{3}(t)-4A_{1}(t)B_{1}(t)\Big)~.
\end{equation}
Finally  substituting back the coefficients $A_{1}(t)$ and $B_{1}(t)$ from Eq. (\ref{H1t}) in terms of the original ones in $A_{0}(t)$ and $B_{0}(t)$, we have 
the norm of the original Hamiltonian of the time-dependent generalised oscillator to be 
\begin{equation}\label{NormHgo1final}
	\begin{split}
F^{2}_{r}\big(\mathbf{H}_{m_1\Omega_1}(t)\big)=\lambda_{1}^{2}f^{2}_{3}(t)+\lambda_{2}^{2}\Big(f^{2}_{3}(t)-4A_{0}(t)B_{0}(t)+C_{0}^{2}(t)\Big)~\\
 = \lambda_{12}^2 f^{2}_{3}(t) - \lambda_{2}^{2} \Big(4A_{0}(t)B_{0}(t)-C_{0}^{2}(t)\Big) ~, 
\end{split}
\end{equation}
where the function $f_3(t)$, defined in Eq. \eqref{fs} above,  can be written in terms of the original time-dependent coefficients as 
\begin{equation}
	f_3(t)=\frac{\dot{\rho}^{2}}{B_{0}(t)}+\Big(A_{0}(t)-\frac{C_{0}^{2}(t)}{4B_{0}(t)}\Big)\rho^2+\frac{B_{0}(t)}{\rho^{2}}~.
\end{equation}
In the second expression for the cost function above, we have written it in a form that is similar to the cost of OTF in Eq. \eqref{NormH5final}. 
If we assume that the time-dependent coefficients in the Hamiltonian \eqref{H1t} are such that the frequency of OTMF is always a positive function
of time, and $B_0(t)>0$, then the role of the second term is to reduce the cost function. 
In Appendix \ref{ckosc} we provide  an example of computation of this cost function for given time-dependent profiles of mass of frequency of the oscillator. 

In this case we can derive a bound on the total cost by following a similar procedure as in section \ref{com-geo} to be 
\begin{equation}\label{bound2}
	\mathcal{C}_{m_1\Omega_1}(T)~\leq~\lambda_{12}T \sqrt{\overline{f_3^{2}(t)}}~,
\end{equation}
where $T$ is final time up to which the total cost has been calculated, and $\overline{f_3^2(t)}$ denotes time average of $f_3^2(t)$ up to the final time.

\subsection{ $su(1,1)$ algebra and the time evolution operator}

We now explicitly consider the time-evolution operator associated with the OTMF, which is essentially
the  solution given in Eq. (\ref{pathordered}) of the corresponding Schrodinger equation (\ref{Hamiltonian}), with the path order replaced by the usual time ordering. In many cases, a compact expression for the time evolution operator  ($\mathcal{U}(t)$) can be derived with the help of the Lie  group theory \cite{Dattoli:1988zza, CFL}. Below, we see that the Hamiltonian  of OTMF is one such example, and thus, it is possible to obtain  a compact  expression for the corresponding time evolution operator and study
various aspects of its associated bi-invariant cost.  

First we notice that the Hamiltonian in  (\ref{H1t}) is an element of the $su(1,1)$ Lie algebra. This can be easily seen by introducing the following  operators  
written in terms of the position and the momentum operators as
\begin{equation}\label{ktoxp}
	K_{+}=\frac{i}{2}\mathbf{X}^{2}~,~~K_{0}=\frac{i}{4}\big(\mathbf{X}\mathbf{P}+\mathbf{P}\mathbf{X}\big)~,~~K_{-}=\frac{i}{2}\mathbf{P}^{2}~,
\end{equation}
and noticing that Hamiltonian of an OTMF (in  \eqref{H1t}), with $m_{1}(t)=1/B_{1}(t)$ and $\Omega_1(t)=\sqrt{A_{1}(t)B_{1}(t)}$,  can be rewritten as 
\begin{equation}
	\begin{split}
\mathbf{H}_{m_1\Omega_1}(t)=a_{+}(t)K_{+}+a_{0}(t)K_{0}+a_{-}(t)K_{-}~,~~\text{with}~\\ 
~a_{+}(t)=-im_{1}(t)\Omega_{1}(t)^{2}~,~~ a_{0}(t)=0~, ~~\text{and}~~a_{-}(t)=\frac{-i}{m_{1}(t)}~~.
\end{split}
\end{equation}
By using the commutation relation between the position and momentum operators, it is easy to check that  $K_{i}$ are the generators of the $su(1,1)$
Lie algebra, and they satisfy the usual commutation relations 
\begin{equation}\label{su11}
	\big[K_{+},K_{-}\big]=-2K_{0}~~, ~~ \big[K_{0},K_{\pm}\big]=\pm K_{\pm}~.
\end{equation}

The fact that the Hamiltonian is an element of the $su(1,1)$ algebra naturally indicates that the  corresponding time evolution operator is an element of the $SU(1,1)$ group. Hence, it can be written as the following two equivalent forms 
\begin{equation}\label{UOTMF}
\mathcal{U}(t,0)=\exp\Big[b_{0}(t)K_{0}+b_{+}(t)K_{+}+b_{-}(t)K_{-}\Big]\equiv \exp\Big[c_{+}(t)K_{+}\Big]\exp\Big[c_{0}(t)K_{0}\Big]\exp\Big[c_{-}(t)K_{-}\Big]~,
\end{equation}
where $b_i$ and $c_i$ are functions of time. It is possible to determine the relationship between these two classes of functions by using the 
so-called decomposition formulas,  and are given by the following \cite{Ban}
\begin{equation}\label{relbc}
	\begin{split}
c_0(t)=-2\ln\big[g(\chi)\big]~,~~c_\pm=\frac{b_\pm \sin \chi}{\chi g(\chi)}~, ~\text{where}~ ~\\
 g(\chi)=\cos \chi -\frac{b_0}{2\chi}\sin \chi~,~\text{and}~~\chi^2=b_+b_--\frac{b_0^2}{4}~.
\end{split}
\end{equation}
These relations can also be established by using a $2\times2$ matrix representation of the corresponding group element.\footnote{See Appendix \ref{decomp} 
for a brief  derivation of these decomposition formulas using the matrix representations. } 
Given  the time dependence of the mass and angular frequency of the OTMF, 
we can determine the functions $c_i$ by using the  following relations \cite{CFL}
\begin{equation}\label{c's}
\begin{split}
c_{+}(t)=m_{1}(t)\frac{d \ln x_{c}(t) }{d t}~,\quad\text{with}\quad c_{+}(t=0)=0~,\\
c_{0}(t)=-2\ln\bigg|\frac{x_c(t)}{x_c(0)}\bigg|~,\quad\text{and}\quad c_{-}(t)=-x_c^{2}(0) \int_{0}^{t} \text{d}t^{\prime}\frac{1}{m_{1}(t^{\prime})x_c^{2}(t^{\prime})}~,
\end{split}
\end{equation}
where the function $x_{c}(t)$ is a solution of the classical equation of motion for the OTMF
\begin{equation}
\frac{d^{2}x_c(t)}{dt^2}+\xi(t)\frac{dx_c(t)}{dt}+\Omega_{1}^2(t)x_c^2(t)=0~,\quad\text{with}\quad\xi(t)=\frac{d\ln m_1(t)}{dt}~,
\end{equation}
and $x_c(0)$ denoting the value of $x_c$ at the initial time $t=0$. 
The condition at $t=0$, i.e., $c_{+}(t=0)=0$   is chosen in such a way that the time evolution operator  satisfies the criterion $\mathcal{U}(0,0)=1$. Given a solution to this equation, using the above relations, we can obtain the required functions  $c_i$s and hence the time evolution operator.\footnote{The corresponding evolution operator of the OTF can be obtained by using an entirely similar procedure, with replacing $m_1=\text{constant}$ and $\Omega_{1}=\Omega$
	in the relations in Eq. \eqref{c's}.} Also, one can, in principle, invert the relations given in Eq. (\ref{relbc}) to obtain the functions $b_i$s as well. However,
in most cases, these are rather  difficult to invert.

\subsection{Equivalence between  OTF  and  OTMF under time reparametrisation}
We  now discuss an interesting equivalence between the OTF (whose  Hamiltonian $\mathbf{H}_{\Omega}(\tau)$ is given in Eq. (\ref{OTF}))  and  OTMF (having the Hamiltonian $\mathbf{H}_{m_1\Omega_1}(t)$  written in Eq. (\ref{H1t})) in terms of the  quantities discussed so far (such as the auxiliary differential equations, the functions $f(t)$ and $f_{3}(t)$ appearing in the expressions for the cost, etc).  For clarity,  we have denoted the time parameters in the Hamiltonian to be different in two cases ($\tau$ and $t$, for OTF and OTMF, respectively), and furthermore, we also assign the corresponding time evolution operators  different symbols,   $\mathcal{\tilde{U}}(\tau)$ for OTF and $\mathcal{U}(t)$ for OTMF, respectively. As we shall see,  these two unitary operators map to each other under suitable mapping between the respective time parameters $\tau$ and $t$. In the next 
subsection we shall discuss  the importance of this equivalence for  calculating  
total costs associated with these time evolution operators. 

To proceed further, we first note that the Schrodinger equation for an OTMF with mass $m_1(t)$ and frequency $\Omega_1(t)$ can be mapped to that of an OTF with unit mass and time-dependent frequency $\Omega(\tau(t))=m_1(t)\Omega_1(t)$ by redefining the time parameter as
\begin{equation}\label{repara}
\text{d}\tau=B_1(t) \text{d}t=\frac{\text{d}t}{m_1(t)}~~\rightarrow~~\tau=\int \frac{\text{d}t}{m_1(t)} +c_3~,
\end{equation}
and vice versa.  Here, $c_3$ is an integration constant. 
In the following, we  assume that the  integral above  can be performed for the range of the parameter $t$  of our interest and the resulting expression for 
$\tau$ can be inverted to obtain the function $t(\tau)$. This is certainly  the case for  the examples we have discuss in this paper.  For example,  
for the Caldirola–Kanai  oscillator considered in the Appendix.  \ref{ckosc},  these two parameters are related by the formula
\begin{equation}
	\tau=-\frac{1}{\delta M}\exp\big[-\delta t\big]+c_3~\quad \rightarrow \quad t=-\frac{1}{\delta}\ln [\delta M(c_3-\tau)]~.
\end{equation}
A change of the time scale of the form \eqref{repara} can be thought of as type of generalised canonical transformation (see e.g., \cite{Leach} for a discussion). 

We  now check the effect of this transformation on other quantities considered so far in this paper.  
First consider the  behaviour of the auxiliary equation Eq. (\ref{auxiliary2}) under the transformation  \eqref{repara}. It can be easily checked that, since $B_1(t)\neq 0$,  under the 
reparametrisation (\ref{repara}) the auxiliary equation  reduces to the following 
\begin{equation}
\frac{d^{2}\rho(\tau)}{d\tau^{2}}+\frac{A_{1}(\tau)}{B_{1}(\tau)}\rho(\tau)-\frac{1}{\rho(\tau)^{3}}=0~,
\end{equation}
and this subsequently  can be recognised as the auxiliary differential equation (Eq. (\ref{auxiliary})) corresponding to that of an  OTF with mass and  frequency, respectively,
 given by 
\begin{equation}\label{map}
 m=1~, ~~~~\text{and}~~~	\Omega(\tau(t))=m_{1}(t)\Omega_{1}(t)=\sqrt{\frac{A_{1}(\tau)}{B_{1}(\tau)}}~.
\end{equation}
Similarity, the function $f_3(t)$, which appears in the expression for the norm  transforms as 
\begin{equation}\label{ff3}
f_3(t)=\frac{\dot{\rho}^{2}}{B_{1}(t)}+A_{1}(t)\rho^2+\frac{B_{1}(t)}{\rho^{2}}\rightarrow~B_{1}(t(\tau))\Bigg[\Bigg(\frac{d\rho(\tau)}{d \tau}\Bigg)^{2}+\frac{A_{1}(\tau)}{B_{1}(\tau)}\rho(\tau)^2+\frac{1}{\rho(\tau)^{2}}\Bigg]=B_{1}(\tau)f(\tau)\Bigg|_{\Omega(\tau)=\sqrt{\frac{A_{1}(\tau)}{B_{1}(\tau)}}}~.
\end{equation} 
As indicated by the last expression, we have identified the expression inside the square bracket  as the function $f(\tau)$ defined in Eq. (\ref{diagonal1}) for OTF with frequency given in Eq. \eqref{map}.

By using the Lewis- Riesenfeld theory of invariant operators, the wave function of both the systems OTF and OTMF can be obtained in terms of the corresponding auxiliary function $\rho$ and its derivative (see, for example, \cite{IAP,CK}). Using these wavefunctions it can be easily checked that under  change of the 
time parameter in  Eq. (\ref{repara}), they map to each other when $\Omega(t)$ is taken to be equal to $m_{1}(t)\Omega_{1}(t)=\sqrt{A_{1}(t)B_{1}^{-1}(t)}$.

The above results clearly indicate that, as far as different quantities of interest are concerned, 
 under the transformation in Eq. (\ref{repara}) the OTMF (with Hamiltonian $\mathbf{H}_{m_1\Omega_1}(t)$) 
having mass $m_1(t)$ and frequency $\Omega_1(t)$ maps to a unit mass OTF ($\mathbf{H}_{\Omega}(\tau)$) whose frequency is related to 
$m_1(t)$ and $\Omega(t)$ through the relation in \eqref{map}. 

Finally, we establish equivalence between the time evolution operators generated by the Hamiltonians of  the two systems. The time evolution operator generated by OTMF is given by  Eq. (\ref{UOTMF}),  with the expressions for the functions appearing  in that form  given in Eq. (\ref{c's}). Under the reparametrization 
in Eq. (\ref{repara}), these functions  transform  to
\begin{equation}\label{c's2}
\begin{split}
c_{+}(\tau)=\frac{d \ln x_{c}(\tau) }{d \tau}~,~~\text{with}~~~c_{+}(\tau_0)=0~,\\
~c_{0}(\tau)=-2\ln\bigg|\frac{x_c(\tau)}{x_c(\tau_0)}\bigg|~,~\text{and}~~ 
c_{-}(\tau)=-x_c^{2}(0) \int_{\tau_0}^{\tau_1}\text{d}\tau^{\prime}\frac{1}{x_c^{2}(\tau^{\prime})}~,
\end{split}
\end{equation}
which are just those of an OTF having unit mass.  Similarly, the differential equation satisfied by the function $x_c(t)$,
\begin{equation}
\frac{d^{2}x_c(\tau)}{d\tau^2}+m_{1}^{2}(\tau)\Omega_{1}^2(\tau)x_c^2(\tau)=0~,
\end{equation}
is just the classical equation of motion of an OTF with $\Omega(\tau)=m_{1}(\tau)\Omega_{1}(\tau)$.  Here $\tau_0$ and $\tau_1$,  respectively, are the values of $\tau(t)$ at the points $t=0$ and $t=1$. Thus, the classical equation of motion of an OTMF with mass $m_1(t)$ and frequency  $\Omega_1(t)$ reduces to the equation of motion of the OTF with unit mass and time-dependent frequency $\Omega(\tau)=m_1\Omega_1(t)$ under the same time reparametrisation as before. Hence, the time evolution operator in Eq. (\ref{UOTMF}) transforms to 
\begin{equation}\label{UOTF}
\mathcal{\tilde{U}}(\tau,\tau_0)= \exp\big[c_{+}(\tau)K_{+}\big]\exp\big[c_{0}(\tau)K_{0}\big]\exp\big[c_{-}(\tau)K_{-}\big]~,
\end{equation}
which is just the time evolution operator generated by an OTF Hamiltonian. This completes our demonstration that under the reparametrisation of time two systems,
an OTMF and an OTF are equivalent.

\subsection{Equivalence between the  cost functional of the time evolution   operators for  the OTF  and  OTMF }
Having established that under the change of time parameter given by Eq. (\ref{repara}), different  quantities associated with  two systems under consideration 
map to each other,  
we now derive a similar equivalence between the total bi-invariant  cost associated with operators generated by  the two Hamiltonians.  Let us first write down the corresponding cost functional evaluated along the given curve  $\mathcal{U}(t)$ (generated by $\mathbf{H}_{m_1\Omega_1}(t)$), which, as we have just shown,  maps to  $\mathcal{\tilde{U}}(\tau)$ (generated by $\mathbf{H}_{\Omega}(\tau)$)  under the said transformation.
For an OTF,  from the expression for the cost function derived in Eq. (\ref{NormH5final}), we see that the total cost of preparing an operator $\tilde{O}$ starting from the identity $\mathbf{I}$ along a curve $\mathcal{\tilde{U}}(\tau)$ is given by 
\begin{equation}\label{costH5}
\begin{split}
\mathcal{D}_\Omega\big(\tilde{O}\big)= \int_{0	}^{1}F\big( \mathcal{\tilde{U}}(\tau),~\mathbf{H}_{\Omega}(\tau)\big) \text{d}\tau~=~\int_{0}^{1}\sqrt{\lambda_{1}^{2}f^{2}(\tau)+\lambda_{2}^{2}\Big(f^{2}(\tau)-4\Omega^{2}(\tau)\Big)} ~\text{d}\tau~\\
~\text{with}~~\mathcal{\tilde{U}}(\tau=0)=\mathbf{I}~,~\text{and}~~\mathcal{\tilde{U}}(\tau=1)=\tilde{O}~,
\end{split}
\end{equation}
while  for an OTMF, using Eq. \eqref{NormH1final}  the  total cost of creating an operator $\mathcal{O}$ starting from the identity $\mathbf{I}$  along the curve $\mathcal{U}(t)$  can be similarly written as 
\begin{equation}\label{costHmo}
\begin{split}
\mathcal{D}_{m_1\Omega_1}\big(O\big)= \int_{0	}^{1}F\big(\mathcal{U}(t),~\mathbf{H}_{m_1\Omega_1}(t)\big) \text{d}t~=\int_{0}^{1}\sqrt{\lambda_{1}^{2}f^{2}_{3}(t)+\lambda_{2}^{2}\Big(f^{2}_{3}(t)-4A_{1}(t)B_{1}(t)\Big)} ~\text{d}t~\\
\text{with}~\mathcal{U}(t=0)=\mathbf{I}~,~\text{and}~~\mathcal{U}(t=1)=O.
\end{split}
\end{equation}
The explicit form of the operators $\mathcal{O}$ and $\mathcal{\tilde{O}}$ can be obtained by evaluating the expressions for the time evolution operators given above in Eqs. (\ref{UOTMF}) and (\ref{UOTF})  at $t=1$ and $\tau=1$ respectively.

We now apply the change of parameter in Eq. (\ref{repara}) to the expression for the  cost in Eq. (\ref{costHmo}). By a little manipulation using Eq. (\ref{ff3}) and finally comparing with Eq. (\ref{costH5}) we can rewrite it in the following suggestive form 
\begin{equation}\label{costHmoho}
\begin{split}
\mathcal{D}_{m_1\Omega_1}\big(O\big)=\int_{\tau_0}^{\tau_1}\sqrt{\lambda_{1}^{2}f^{2}(\tau)+\lambda_{2}^{2}\Big(f^{2}(\tau)-4\frac{A_{1}(\tau)}{B_{1}(\tau)}\Big)}~ \text{d}\tau~=\mathcal{D}_{\Omega}\big(O\big)\\
	\text{with}~\mathcal{U}(t=0)\equiv\mathcal{U}(\tau_0)=\mathbf{I}~,~\text{and}~~\mathcal{U}(t=1)\equiv\mathcal{U}(\tau_1)=O.
\end{split}
\end{equation}
Here,  as before, $\tau_0$ and $\tau_1$ respectively, are the values of $\tau(t)$ at the points $t=0$ and $t=1$, and we have used the fact that the time evolution operators of the two systems maps to each other under the said reparametrisation. In general, the values of $\tau(t)$ do not coincide with those of $t$ at the two points of the given unitary. However, by choosing the integration constant $c_3$ in Eq. \eqref{repara}, we can initialise both to be  zero at the starting point of the given unitary curve. This essentially means that at $t=0$ and $\tau=0$, the corresponding unitary time evolution operators are unity - a condition automatically satisfied by the evolution operator with the expressions given in the previous section. 

The relation  derived in \eqref{costHmoho} indicates that the total cost of creating an operator $O$  evaluated along the given curve $\mathcal{U}(t(\tau))$,  when we consider it to be generated by the Hamiltonian $\mathbf{H}_{m_1\Omega_1}(t)$,  is equal to the total cost when it is generated by the Hamiltonian  $\mathbf{H}_{\Omega}(\tau)$, with the relations between the time-dependent parameters of these two Hamiltonian are the same as the one we have derived  in Eq. \eqref{map} from the mapping between different quantities associated with the two systems.

In view of this relation between the total costs,  here we discuss some important points which should   further clarify the meaning 
and the implications of this equivalence. 

\textbf{1.} This equivalence between the cost functionals  implies that the total cost of creating a given unitary operator $O$ starting from the identity does not depend on which of the two Hamiltonians (and also the respective parametrisation) we choose to use. 
Furthermore, it is to be kept in mind that in both the cases, the cost has to be calculated between the same two points along the path, i.e. between
identity and the operator $O$ under considerations. Suppose we known the unitary time evolution curve $\mathcal{U}(t)$  generated by Hamiltonian $\mathbf{H}_{m_1\Omega_1}(t)$ (as we have just  seen this can be obtained using Lie algebraic methods), then we can obtain the curve $\mathcal{\tilde{U}}(\tau)$  generated by $\mathbf{H}_{\Omega}(\tau)$ as well. Since $\tau=1$, in general, does not coincide with the point $t=1$,  the point $\tau=1$ corresponds to a  different operator (say $O_1$) on the unitary curve than the operator we want to create, i.e. $O$.  As mentioned before,  even if we choose the constant $c_{3}$ in such a way that the initial point of the curve corresponds to $t=\tau=0$,  one must still be careful to distinguish the problem of calculating the cost of the operator $O$ from that of creating  $O_{1}$. 

\textbf{2.} It is also important to notice that the relation between the total costs we have established  is not an artifact of the unitary invariance of the cost function. The total costs are equivalent even if the two Hamiltonians are not related by any time-independent unitary transformation. In fact,
the two Hamiltonians in question (an OTF and OTMF) are not related by any time-independent unitary transformation. The equivalence of costs directly follows from the definitions of the time evolution operator itself. During the calculation of the total cost of creating a certain operator generated by a given
 Hamiltonian  we can always perform such a reparametrisation of $t$, however,  unlike the `special' transformation given in Eq. (\ref{repara}), the transformed  total norm usually does not
have any physical meaning as the integral pf the norm of the Hamiltonian  operator  of certain other quantum mechanical system. 
Furthermore, it will be interesting to find out whether such mapping of costs can be found for other pairs of generators, i.e. whether it 
is possible to find out other pairs of `(total) cost-equivalent' generators such that other quantities (e.g., wavefunction or the time evolution operator) 
also map to each other under the same reparametrisation which makes total costs equivalent to each other.  

\textbf{3.} Before moving on,  we observe two more points related to this  mapping between two systems. Firstly, the norms of the two Hamiltonians 
do not map to each other under the said transformation; it is only their integrals, i.e. the cost functional or the total cost of creating a certain operator
which   map to each other. The norms of the  Hamiltonians themselves can be quite different.  In this context, we also want to point out the importance of the relation in Eq. \eqref{ff3} : The presence of the extra factor of $B_1$ in this relation cancels out the $1/B_1$ factor from the time reparametrisation in the total cost so that the equivalence between the cost functional holds.    In fact, this is also the reason why the upper
bounds on the total cost functions in eqs. \eqref{bound} and \eqref{bound2} do not map to each other under time reparametrisation.
 
Secondly,   the Hamiltonians $\mathbf{H}_{m_1\Omega_1}(t)$  in Eq. (\ref{OTF})  and $\mathbf{H}_{\Omega}(\tau)$ in Eq. (\ref{H1t}) can be formally connected to each other through suitable unitary transformations (see e.g. \cite{Mostafazadeh:1998hq} where such transformations were derived).  However, these unitary transformations 
are time-dependent, and therefore, the bi-invariant costs of these two Hamiltonians are not equivalent. 
 

\subsection{Cost and complexity along the time evolution curve generated by  time-dependent Hamiltonians}
The  central problem considered in this paper, namely,  establishing  the CG  of operators generated by the time-dependent Hamiltonian of a quantum mechanical system,    the instantaneous Hamiltonian (and hence the unitary evolution along which the cost has to be evaluated) is the quantity which  we assume to be given. 
This is in contrast with the standard problem of calculating the circuit complexity of some specified operator starting from the identity (or a target state wavefunction from a reference state wavefunction), where the instantaneous generator or the specific unitary curve is not specified beforehand. In that case, one needs to find the
generator and minimize the cost  among all the possible curves connecting two points on the unitary to calculate the complexity. However, here, we only want to 
calculate the total cost  between two points of the specified curve generated by the given time-dependent Hamiltonian, since only this is the meaningful quantity in this case. 
 
It may be possible that the two points in question are connected by many other curves on the space of unitaries and computational cost along some of those are smaller than the ones computed along the given unitary. However, here we are only interested in the cost alone, not the minimum of such costs, and,
furthermore, by our assumption, the given unitary is not geodesic on the bi-invariant CG since the Hamiltonian generator of this curve is not constant.
In fact,  the minimum  cost i.e.,  the complexity, is obtained by evaluating the cost along 
the geodesic, which, according to the bi-invariance,  is generated by a constant Hamiltonian. 

For the time evolution operators considered in this and the previous section, we can actually find out the constant  generator $H_0$ corresponding to the geodesic from the 
expression of an operator $\mathcal{O}$ at a given  point of the evolution curve (say, corresponding to the point $t=1$).  
For an OTMF, from the expression for the time evolution operator given in eq. (\ref{UOTMF}) we see that 
\begin{equation}
	\mathcal{O}=\mathcal{U}(1,0)=\exp\Big[b_{0}(1)K_{0}+b_{+}(1)K_{+}+b_{-}(1)K_{-}\Big]~,
\end{equation}
hence, the constant generator which generates the geodesic is given simply  by 
\begin{equation}\label{contH}
	H_0=i\Big[b_{0}(1)K_{0}+b_{+}(1)K_{+}+b_{-}(1)K_{-}\Big]~.
\end{equation}
This is just the Hamiltonian of a generalised oscillator  with constant coefficients considered in section \ref{time-independent}, where the values of these
coefficients can be read by comparing the above expression and the Hamiltonian in Eq. \eqref{generalosc} and using Eq. \eqref{ktoxp}
for the generators of the $su(1,1)$ Lie algebra in terms of the position and momentum.  We have already considered the 
bi-invariant CG of the operators generated by this type of Hamiltonian in section \ref{time-independent}, and the expression for the cost and complexity can be read off from Eqs. \eqref{GHOcost} and \eqref{BIcomp} respectively.

To summarise, according to the discussions in the previous two subsections,  given a unitary curve generated by the Hamiltonian of the OTMF, we can calculate the total cost of creating an operator  on this curve from the bi-invariant CG. Furthermore,  this cost is the same as that of an operator on the time evolution curve generated by the Hamiltonian of the OTF whose frequency is related to the mass and frequency of the OTMF. However,  since the specified curve is not a  geodesic of the bi-invariant geometry,  this cost is not the minimum one, i.e., it does not correspond to the complexity of the given operator. In this case, the minimum cost is actually along a curve whose generator is the Hamiltonian of a generalised oscillator (as in Eq. \eqref{contH}). 

Before concluding this section we note the following: Given a particular operator at an arbitrary point on the time evolution curve, 
it would be interesting to define an efficiency associated with the evolution, following, e.g.,  \cite{Anandan}, as the ratio of the minimum 
(geodesic) distance computed along the path connecting the identity and the final operator (or the initial and the final states),  and the  length of the actual path generated by the time-dependent Hamiltonian under consideration. The difference between the efficiency introduced  in \cite{Anandan} and our proposal is that the former uses the Fubini-Study 
metric defined on the space pure quantum state to measure the distance,\footnote{See \cite{Cafaro} for a recent work that uses this definition to measure the efficiency of the 
	analog quantum search algorithms.} while our definition uses the bi-invariant CG to measure the same.  It will be interesting  to study this quantity further in a future work. 

\section{Quantum quenches - an exact example of time-dependent protocols}\label{quench}
As a concrete example of the application  of the formula for the  cost function and the associated total cost derived in section \ref{com-geo},
here we discuss  a quantum quench scenario. 
Consider the following profile for the frequency of an OTF, which has been recently proposed in \cite{KA} :
\begin{equation}
	\begin{split}
		\Omega(t)=\Omega_{0}=\frac{\delta}{\eta^{2}} \quad \text{for} \quad t\leq0~,\\
		=\frac{\delta}{t^{2}+\eta^{2}} \quad \text{for}  \quad t>0~.
	\end{split}
\end{equation}
Here, $\eta$ and $\delta$ are two constants. 
The system  therefore is described by the usual HO up to a time $t_0=0$,  after which the frequency of the oscillator is changed continuously 
with time from a maximum value at $t=0$. In  the limit of infinite time,  the time-dependent  frequency approaches to  zero. Quantum quenches where the  frequency 
exhibits this kind of time dependence are known as the end-critical protocol in the literature.
For this particular  form of the frequency, the exact solution of the auxiliary equation (\ref{auxiliary}) is known and can be written in terms of elementary functions.
The evolution of  different entanglement measures such as the Shannon entropy, Entanglement entropy,  out-of-time order correlator, 
different measures of circuit complexity were studied in such models (an OTF whose
frequency starts changing with time after some initial time) by different authors, see, e.g., \cite{Chandran:2022vrw, KA, Pal:2022rqq, Ghosh:2017nlk,DiGiulio:2021oal, Camargo:2018eof} for a selection of references. 

To find out  the values of the constants appearing in the expression for $\rho(t)$ we use the following initial conditions,  
\begin{equation}\label{t=0conditions}
	\rho(t=0)=\frac{1}{\sqrt{\Omega_{0}}}~, \quad \text{and} \quad \dot{\rho}(t_0=0)=0~.
\end{equation}
These conditions  ensure that at the initial time $t_0=0$, just before the start of the quench, 
the wavefunctions of the OTF match with those of the eigenstate of the instantaneous Hamiltonian $\mathbf{H}_{\Omega}(t_0=0)$.
The solution of the auxiliary equation for the entire range of time, which satisfies the conditions listed above,  is given by
\begin{equation}
	\begin{split}
		\rho(t)=\frac{\eta}{\sqrt{\delta}} \quad \text{for} \quad t\leq0~,\\
		=\frac{\eta}{\sqrt{\delta}} \bigg[\frac{t^{2}+\eta^{2}}{\delta^{2}+\eta^{2}}\bigg(\cos^{2}\Big[\lambda \arctan \Big(\frac{t}{\eta}\Big)\Big]+\frac{\delta^{2}}{\eta^{2}}\bigg)\bigg]^{1/2} \quad \text{for} \quad t> 0~, 
	\end{split}
\end{equation}
where $\lambda=\big(1+\frac{\delta^2}{\eta^2}\big)$.  As can be easily checked, by using the well-known solutions of the time-dependent 
Schrodinger equations for OTF \cite{LR},  the initial conditions ensure that these wavefunctions match with the eigenfunctions of a HO
with frequency $\delta/\eta^2$ at the time $t=0$. 

Here our  goal is to find out the cost function and the total cost of the time evolution operator and compare it with other well-known measures of quantum information 
such as the Shannon entropy (SE).

\subsection{Change in the Shannon entropy}
Let us first discuss the evolution of the SE for this profile in this quench protocol.  
For the $n$-th state wavefunction $\big|\psi_{n}(t)\big>$, one can show, by using  the standard formula for the SE,  it actually depends on $n$ \cite{CKKMMN}. 
However, as recently observed in \cite{KA},  the difference of SE before and after the start of the quench at $t=0$  does not depend on $n$ and is given by the simple formula 
\begin{equation}\label{delS}
	\Delta S_n(t)=S_{n}(t)-S_{n}(t_0)=\ln \big(\sqrt{\Omega_{0}}\rho(t)\big)~.
\end{equation}
We study the properties of this quantity in various cases, such as slow and fast quench scenarios and early and late times. First we introduce the dimensionless time parameter $s=t/\eta$, and furthermore,  denote $\delta=\alpha\eta^{2}$ and $\beta=\alpha\eta$. Then the solution for $\rho(t)$ for $t>0$ can be rewritten as 
\begin{equation}\label{rhos}
	\rho(t)=\sqrt{\frac{\eta}{\beta} \bigg(\frac{s^{2}+1}{1+\beta^{2}}\bigg)} \bigg(\cos^{2}\Big[\sqrt{1+\beta^{2}} \arctan (s)\Big]+\beta^{2}\bigg)^{1/2}~.
\end{equation} 

We now  consider two different  cases,  respectively $\beta \gg 1$,  and $\beta \ll 1$  at late times, and analyse properties of the SE in  these limits.

\textbf{Case-1: $\beta \gg 1$:}
In this limit we first need to find the KZ time ($t_{KZ}$) where the adiabaticity breaks down. This is determined by the Landau criterion $\frac{|\dot{\Omega}(t)|}{|\Omega^{2}(t)|} \sim 1$.
For the above frequency profile, we can easily determine it to be $t_{KZ}\sim\frac{\alpha \eta^{2}}{2}$, i.e. in terms of dimensionless time $s_{KZ}\sim \beta$. The initial time evolution before this time scale is adiabatic. 

In the adiabatic approximation we then have $\rho(s_{KZ})=\frac{1}{\sqrt{\Omega(s_{KZ})}}$. Thus, the adiabatic approximation predicts the difference between the SE before and after the quench at $t=0$ (from Eq. \eqref{delS}) to be 
\begin{equation}
	\Delta S_n\sim \frac{1}{2}\ln \big(\beta^{2}+1\big)~=\frac{1}{2}\ln \big(s_{KZ}^{2}+1\big)~.
\end{equation}
Since at $s_{KZ}$ the adiabatic approximation breaks down, this result is valid for times earlier than $s_{KZ}$ i.e. for $s<s_{KZ}\sim \beta$. 
For fixed values of $s$ before the KZ time, the change in  SE can be written as 
\begin{equation}\label{slowSE}
	\Delta S_n(s)\sim \frac{1}{2}\ln \big(s^{2}+1\big)\quad \text{for} \quad \beta>s~.
\end{equation}

\begin{figure}[h!]
	\begin{minipage}{0.4\linewidth}
		\centering
		\includegraphics[width=2.9in,height=2.2in]{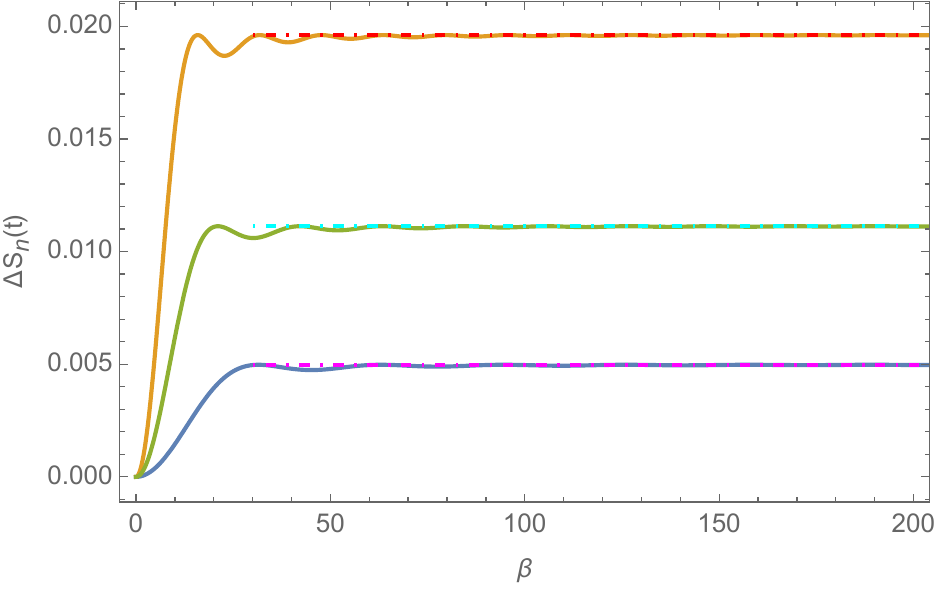}
		\caption{Comparison of exact (solid curves) and approximate expressions for the  change in SE with adiabatic approximation, i.e. before the KZ time 
			(dot-dashed lines), for fixed values of $s$ at early times. 
			Dot-dashed lines: magenta ($s=0.1$), cyan ($s=0.15$)  and red ($s=0.2$). 	
			Solid curves are plotted  with the same values $s$ as their associated approximated counterparts. }
		\label{fig:Shannon-largebeta}
	\end{minipage}
	\hspace{2. cm}
	\begin{minipage}{0.4\linewidth}
		\centering
		\includegraphics[width=2.9in,height=2.2in]{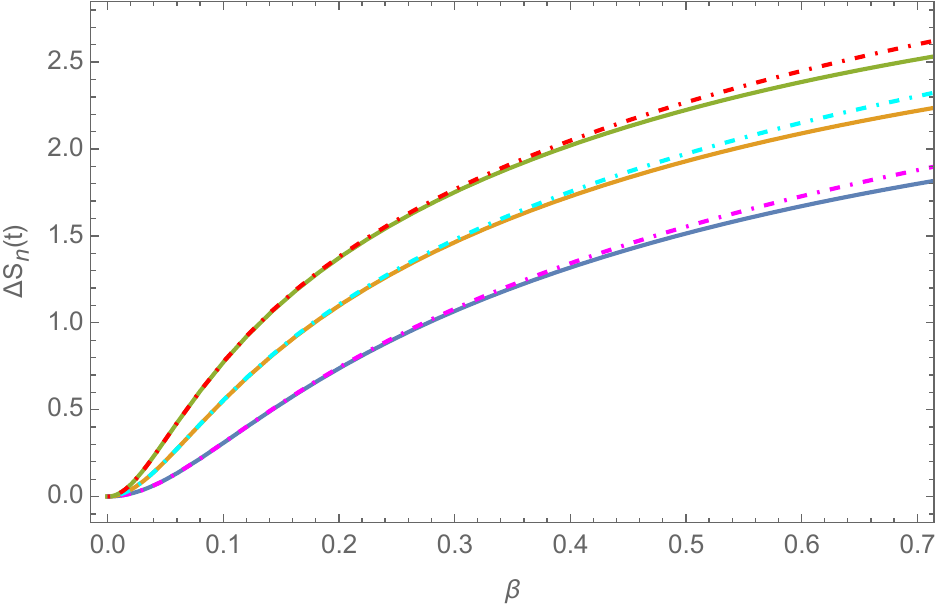}
		\caption{Comparison of exact (solid curves) and approximate expressions (dot-dashed  curves)  of the  change in SE at late times for small $\beta$,
			with fixed values of $s$. Dot-dashed: magenta  ($s=10$), cyan ($s=15$) and red ($s=20$). 
			Solid curves are plotted with the same values $s$ as their associated approximated counterparts. }
		\label{fig:Shannon-smallbeta}
	\end{minipage}
\end{figure}

We graphically compare this approximate formula  with the exact result for fixed values of $s$.
In fig - \ref{fig:Shannon-largebeta}, we have plotted both the exact expression for the change in SE using  Eqs. (\ref{rhos}) and \eqref{delS}
(indicated by solid curves), as well as the one obtained  using the adiabatic approximation (indicated by the dot-dashed lines) for different fixed values of the time parameter $s$. 
The approximate results are  valid up to a time before the KZ time. From the exact result it can be seen that for relatively small values of $\beta$, the change in SE grows and 
then shows oscillatory behavior, however these oscillations dry out quickly, and $\Delta S_n$ takes a fixed constant value for large $\beta$
which can be predicted quite accurately by using  
the adiabatic approximation (the dot-dashed lines merge with the solid ones for large $\beta$).  A comparison of plots with different fixed $s$ shows that with increasing  values of $s$, the oscillations  around the fixed adiabatic value increase and occur more rapidly. 

\textbf{Case-2. late times and $\beta\ll 1$:}
In  the limit $\beta \ll1$, and $t\gg\beta$ i.e. $s\gg1$, expanding the exact expression for  $\rho^{2}$ in Eq. (\ref{rhos}) and neglecting terms of order $\beta^2$ and higher   we get  \cite{KA}
\begin{equation}\label{rholate}
	\rho^{2}(s)\simeq \frac{\eta}{\beta}+\eta\Big(s^2-\frac{s \pi}{2}+1\Big)\beta+\mathcal{O}(1/s)~.
\end{equation}
The corresponding expression for the change in the SE is therefore given by
\begin{equation}\label{fastSE}
	\Delta S_n(t)\sim \frac{1}{2}\ln \Big(1+\Big(s^2-\frac{s \pi}{2}+1\Big)\beta^2\Big) \quad \text{for} \quad \beta \ll1~, ~~\text{and}~~s\gg1~. 
\end{equation}
In fig - \ref{fig:Shannon-smallbeta}, the change in SE is plotted with small $\beta$ and at late times and is compared with the exact expression.  
It can be seen that for sufficiently small $\beta$,  Eq. (\ref{fastSE}) provides quite a good approximation to the change in SE for different times.


\subsection{Cost  of the time evolution operator}
\begin{figure}[h!]
	\begin{minipage}{0.4\linewidth}
		\centering
		\includegraphics[width=3in,height=2.2in]{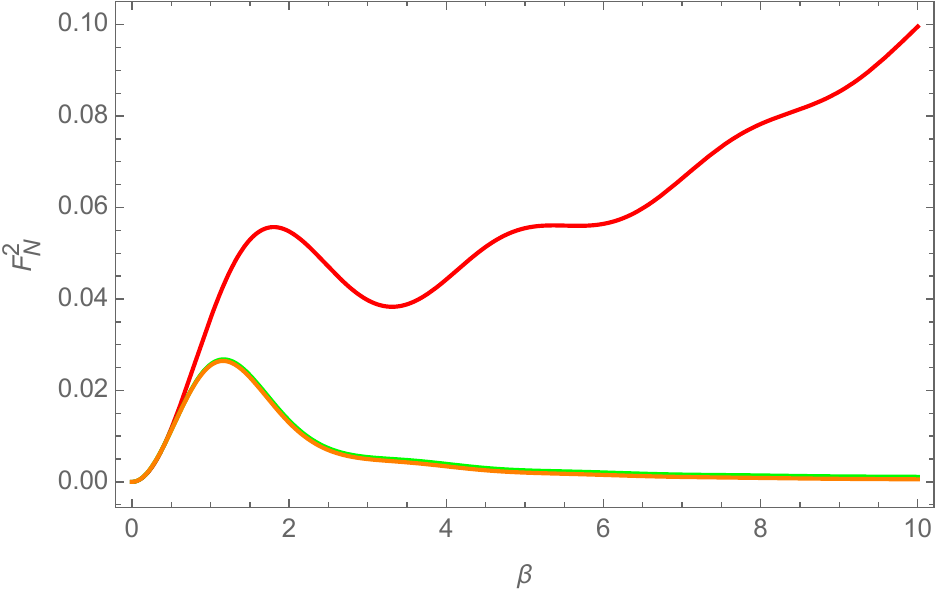}
		\caption{Variation of the cost ($F_N^2 (s)= \eta^2 F_r^2 (s)$) with respect to $\beta$ for different values of $s$. Here the  fixed values of the dimensional time 
			are $s=2$ (red), $s=20$ (green), and $s=200$ (orange). For small values of $\beta$ ($0<\beta<1$) the cost always grows irrespective of the 
			value $s$ and subsequently reaches a local peak. }
		\label{fig:costlatetime}
	\end{minipage}
	\hspace{2. cm}
	\begin{minipage}{0.4\linewidth}
		\centering
		\includegraphics[width=3in,height=2.2in]{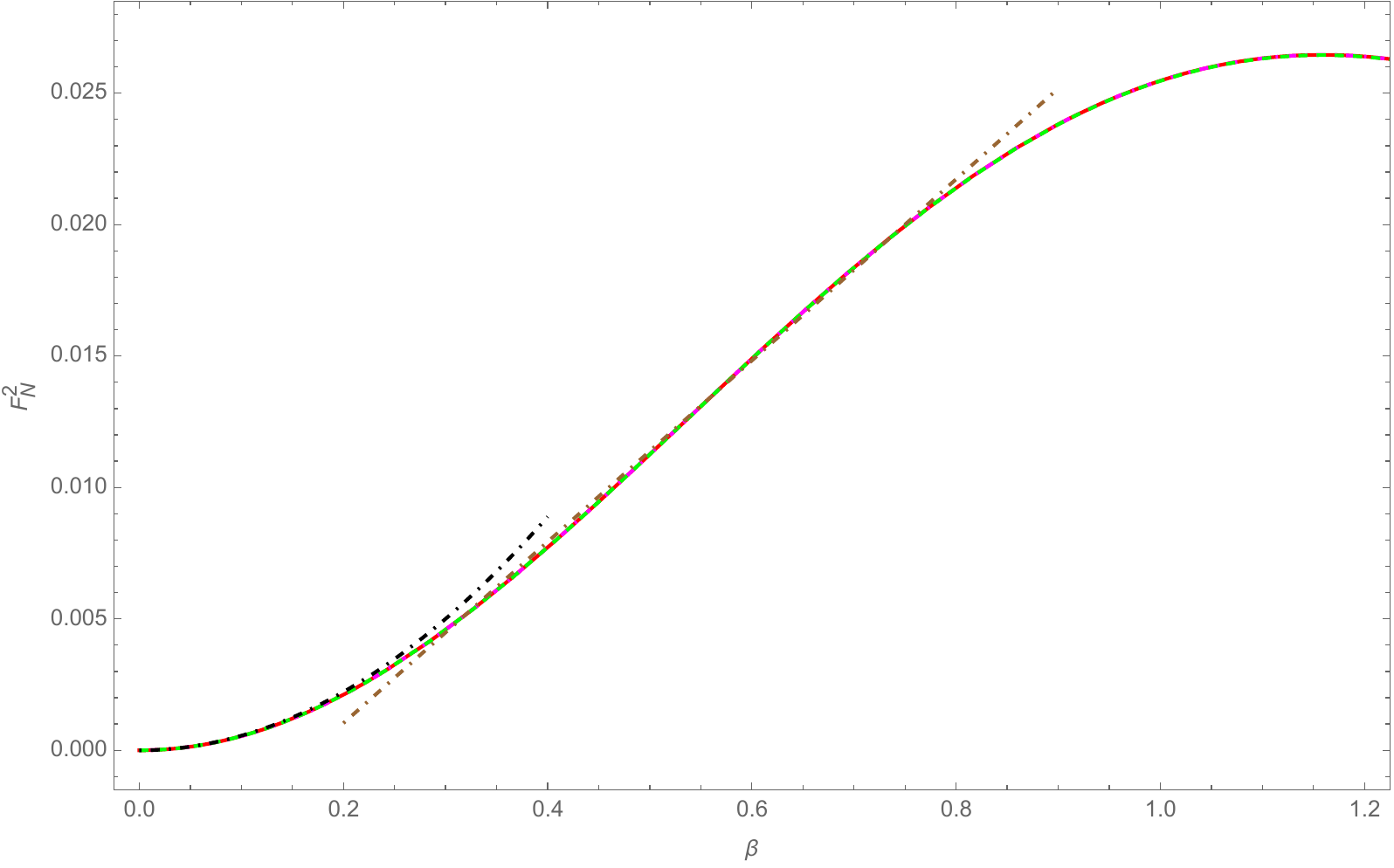}
		\caption{Scaling of the cost ($F_N^2 (s)= \eta^2 F_r^2 (s)$) with respect to $\beta$ at late times for fast quenches. The dashed green curve represents the approximate cost obtained using Eq. \eqref{rholate}, while the solid red is the exact one. Quadratic and linear scaling are also shown by dot-dashed black and brown fittings.  Here, we have  fixed  dimensional time 
			$s=200$. }
		\label{fig:costscaling}
	\end{minipage}
\end{figure}

Next, we  compute the  cost of the time evolution operator for this frequency profile with the help of the formula derived in section \ref{OTFCG} and study its various limits. The expression for the time-dependent function $f(t)$, obtained  by using  the solution of the auxiliary function written above  (see Eq. (\ref{rhos}))
into Eq. \eqref{diagonal1}, is given in terms of the dimensionless time $s$  by
\begin{equation}
	f(s)=\frac{1}{2 \beta \eta (1+s^2) (1+\beta^2)} \Big((1+2 \beta^2) (1+2\beta^2+s^2) + (s^2-1) \cos (\Theta(s))- 2s \sqrt{1+\beta^2} \sin (\Theta(s))\Big)~,
\end{equation}
where we have defined $\Theta(s) = 2\sqrt{1+\beta^2} \arctan(s)$. Substituting this expression in Eq. \eqref{NormH5final} we get the expression for the cost. However, this
expression is too cumbersome to reproduce here, and therefore we resort to approximate and  graphical analysis. 
To understand the behaviour of the cost function as a function of $s$, first consider its expansion with respect to $s$ around the initial time ($s=0$). This, 
upto  $ \mathcal{O} (s^4)$,  can be written as 
\begin{equation}
	F_{N}^{2}(s) \simeq  4\lambda_1^2 \beta^2 - 8\lambda_1^2 \beta^2 s^2 + 4\beta^2 (4\lambda_1^2+\lambda_2^2) s^4 + \mathcal{O} (s^6)~.
\end{equation}
Here we have written down the expansion of $F_N^2 (s)= \eta^2 F_r^2 (s)$, instead of just  $F_r^2 (s)$,  so that the resultant expression is actually independent 
of $\eta$.\footnote{In the following, this is the quantity we refer to as cost.} 
In this expansion, only even powers of the dimensionless time parameter $s$ appear, and for small values of it  $F_{N}^{2}(s)$  grows quadratically with $\beta$. 
In fact, as can be easily checked, for $0<s<1$, the cost is always an increasing function of $\beta$.  For $s>1$, the behaviour of the cost is more non-trivial. For small values of $\beta$, within the range $0<\beta<1$, irrespective of the value of $s$,  the cost grows with $\beta$.  While, for $\beta>1$ the nature of the cost depends on the 
exact value of $s$ as well. Roughly, for large $\beta$, the cost decreases towards zero for large values $s$, while 
for small values of $s$, the cost, after reaching a minimum value, keeps increasing for large $\beta$. Thus, there is a transition 
in the behaviour of the cost as we keep increasing $s$ above $1$.  These features can be viewed from the plots of $F_{N}^{2}(s)$ shown in Fig. \ref{fig:costlatetime}, where we have  
taken $\lambda_1^2 \approx0.004$ and $\lambda_{2}^2 \approx 0.05$ for illustration.  Note that,  even though for the range of $\beta$ shown in this plot, the curve with 
$s=20$ seems to be decreasing at large $\beta$, it would actually increase for sufficiently large $\beta$.  Therefore, unlike the change in the SE, the cost does not saturate to a constant 
value for large $\beta$, irrespective of the value of $s$. 

We can obtain  a scaling behaviour of the cost at late times ($s \gg 1$) and for $\beta \ll 1$ (i.e., for the case-2 considered for the SE above).  Using the expansion
of the auxiliary function given in Eq. \eqref{rholate} for late times and small $\beta$, in the expression for the cost, we can obtain an 
approximate behaviour of $F_{N}^{2}(s)$ in this limit.  This is plotted in Fig. \ref{fig:costscaling}  (the dashed green curve),  along with 
the exact expression (red), and as can be seen, the truncated expression provides quite a 
good approximation for the cost function.  Furthermore, in this plot, we have also shown the initial quadratic scaling of the cost for small $\beta$, which 
subsequently  transforms with a linear scaling. For the  fittings shown in Fig. \ref{fig:costscaling}, the numerically obtained equations for the scalings are: quadratic scaling $F_{N}^{2} = \beta^2/18$ (shown by the dot-dashed black curve), 
and for linear scaling $F_{N}^{2} = 0.0345 (\beta -0.17)$ (shown by dot-dashed brown line).

On the other hand,  to study the behaviour of the cost at late times, and arbitrary $\beta$,  we can expand the cost with respect to $1/s$, keeping $\beta$ fixed,  to get the 
following expression
\begin{equation}
	F_{N}^{2}(s) \simeq  \frac{\lambda_{1}^2+\lambda_{2}^2}{(1+\beta^2)^2} \Big[1+2\beta^2 + \cos \Big(\pi\sqrt{1+\beta^2}\Big)\Big]^2 \Big(\frac{1}{4\beta^2}+ 
	\frac{1}{s^2} \Big)+\mathcal{O}\big(\frac{1}{s^3}\big)~.
\end{equation}
Therefore, with fixed values of $\beta$,  at late times after the quench, the cost always takes constant values, whose magnitude  depends on the fixed $\beta$.  
Also, notice that,  at the start of the quench at
$s=0$,  the cost has the same value as that of a constant frequency HO, as can be seen from the general expression for the cost derived in  Eq. \eqref{NormH5final}.

\section{Summary and discussions}\label{conclusion}
In this paper we have discussed a method of obtaining the regularised bi-invariant CG of the operator generated by the time-dependent Hamiltonian of a quantum mechanical system. 
The procedure we have followed  makes use of a regularisation scheme in which the norm of an infinite dimensional Hamiltonian is obtained by first imposing a 
condition on its trace and subsequently using the resultant mean value ($a$) in the calculation of the cost function.  For time-independent Hamiltonians (which we had 
considered in section \ref{time-independent}), this mean value is also a constant, and therefore, the cost itself is independent of time. However, for  explicitly 
time-dependent Hamiltonians considered in later sections, this mean value is itself a function of time, and as a result,  the computation of the regularised cost is more involved. 
Another feature of the calculation where the computation of the cost of the two types of systems differs is that, for time-independent Hamiltonians, one only needs to know the 
energy eigenvalues to calculate the cost, whereas since a time-dependent Hamiltonian has no stationary eigenstates, both  diagonal as well as non-diagonal
matrix elements of this Hamiltonian in a suitable basis are required for the computation of the cost. 
To this end, we have used the well known Lewis-Riesenfeld  method of time-dependent invariant operators since the eigenfunctions 
of this operator can be directly used efficiently to  find out both the  diagonal and non-diagonal matrix elements of the Hamiltonian under
consideration, and hence the norm can be calculated 
in terms of these matrix elements. This is the strategy we have followed in section \ref{OTFCG} and \ref{OTMFCG} to obtain the regularised norm of time-dependent 
Hamiltonians, where the generic properties of the obtained results are also discussed. 

As emphasised in the main text, we again want to stress the importance of looking at the standard problem of circuit complexity from a slightly different
	viewpoint:  rather than finding  a minimum of all unitary curves connecting the two given operators (which is the standard Nielsen's geometric framework of obtained the circuit complexity), calculating the total cost of evolution along a specified path on a CG  is also important, since the time-dependent Hamiltonian of a quantum mechanical system may itself generate such unitary evolution. In that situation, it is meaningful only to compute the cost as a function of time along that
	specified evolution curve, not the complexity. 
	The assumption of bi-invariance of the CG, and the regularisation procedure used to find out the norm for an infinite dimensional Hilbert space, are two key ingredients of our analysis, and both have strong effects on the final form of the cost function we have derived for different classes of harmonic oscillators. 

Among various examples of explicitly time-dependent systems, in this paper, we have focused on two particular Hamiltonians, which are widely studied in the literature 
due to their importance in modelling various physical phenomena.  These are, respectively, an oscillator of unit mass whose frequency of oscillation explicitly depends on time and an 
oscillator whose both mass and frequency change with time. For convenience here we briefly
	list the main results we have obtained in this paper along with some of their implications:
	1.  For an OTF and an OTMF, using the Lewis-Riesenfeld   method, we have obtained compact analytical expressions for the cost of the time evolution operator generated by these two Hamiltonians (see eqs. \eqref{NormH5final} and \eqref{NormHgo1final}, respectively). 
	2. We have shown there exists  an equivalence between the two separate examples considered 
	so that expression for various quantities of two systems can be mapped to one another by a time reparametrisation, 
	and demonstrated the implications of this duality in terms of the total bi-invariant cost. 
	3. We have obtained an upper  bound on the total cost up to a finite time (see eqs. \eqref{bound} and \eqref{bound2}), which is analogous  to the quantum speed limits associated  with various quantum information-theoretical quantities (usually a quantifier of `distance' between two quantum states). 
	4. One of the most important examples of time-dependent quantum systems, namely a smooth quench of the  Hamiltonian parameters is considered in section \ref{quench}. For a time-dependent frequency oscillator model, using the exact analytical expressions available in the literature, we have studied information theoretic quantities, e.g.,  change of the Shannon entropy,  and compared them with the cost of the time-evolution operator derived here. In both cases, we have shown that different quench regimes and their associated characteristics can be effectively probed by these quantifiers
	by using their scaling relations.

We also now briefly discuss the differences between the cost function we have derived in this paper, and the approach followed in general
	for the case of a time-dependent Hamiltonian, with the existing results in the literature. The approach developed in \cite{Myers1, BSS} to compute the NC in quantum field theories, which is one of the most common approaches followed by many authors,  directly 
	uses the wavefunction of the state which differ from each other in terms of the values of the parameter of the system (e.g., the mass and the frequency for a harmonic oscillator), not time,  and is therefore,  not directly related to the time-evolution of the state. On the other hand, refs.  \cite{BDKP1, BDKP2} studied NC 
	of time-evolution operator. However, the role played by the time parameter in \cite{BDKP1, BDKP2} is quite different from the one 
	in this paper,  since, in the approach followed 
	by these authors,  the parameter that connects the initial and final points on the CG is different from the time. Therefore, in our opinion, it is not very meaningful to directly compare our results for the cost function with the complexity that appeared so far in the literature, simply because, in general,  they refer to different quantities, albeit in the same geometrical framework. 

Before concluding, we mention a few possible future directions. Firstly, in this paper we have mainly considered the generalised oscillator and other Hamiltonians which can be 
connected to it by unitary transformations. Extending our analysis to systems beyond this class, specifically systems that are classically chaotic, can be one important direction that can be pursued.  
Secondly, the bi-invariant CG we have discussed in this paper is invariant under unitary transformations between the generators 
which are independent of the parameter describing the unitary  curve, e.g., as we have discussed in section \ref{HOMFnorm}, 
the unitary transformation $U_1$ in \eqref{U1t}
connecting the Hamiltonians of a generalised oscillator and an OTMF  is time-independent, hence, the cost is the same for both the Hamiltonians. However, if 
we consider  two Hamiltonians which are  related through a time-dependent unitary transformation, the CG of  Eq. \eqref{p=1cost} is no longer invariant under 
such transformation (this is the reason why the costs associated with OTF and OTMF are not the same).  
Determining  an invariant  CG generated by  Hamiltonians related through general time-dependent unitary transformations is an interesting problem, and  we hope to come back to it in a future work.

\begin{center}
	\bf{Acknowledgments}
\end{center}
We sincerely thank Hugo A. Camargo,  Yichao Fu, Viktor Jahnke,   Keun-Young Kim, and Run-Qiu Yang for useful discussions on this and 
other related works, as well as comments on a draft version of the manuscript.
We would also like to thank the organisers of the workshop  “Student Talks in Trending Topics in Theory'' (ST$^4$ -2023), during which parts of the present work were carried out. The work of Kunal Pal is supported by the National Research Foundation of Korea under Grants No. 2017R1A2A2A05001422 and
No. 2020R1A2C2008103. We thank our anonymous referees for their constructive comments and criticisms which helped to improve a draft version of this manuscript.

\appendix

\section{Further examples of bi-invariant complexity geometry with time-independent Hamiltonian} \label{timeindexamples}

\subsection{ Complexity geometry generated by the generalised oscillator Hamiltonian with a constant driving force}\label{driving}
So far we have regularised the trace of a Hamiltonian by introducing a regulator ($a$), and  the regularisation criterion (in \eqref{regularization})
in turn, determines this constant.  However,  we need not always introduce any regular by hand; instead, when an extra constant driving force is present in the system, it can be interpreted as a trace regulator.
To see this explicitly, once again consider the Hamiltonian of  the  generalised oscillator in Eq. (\ref{generalosc}), however, now with the addition of an extra potential term linear in the position coordinate 
\begin{equation}\label{osc+driving}
	\mathbf{H}_{F}\big(\mathbf{X},\mathbf{P}\big)=\frac{A_{0}}{2}\mathbf{X}^{2}+\frac{B_{0}}{2}\mathbf{P}^{2}+\frac{C_{0}}{4}\big(\mathbf{X}\mathbf{P}+\mathbf{P}\mathbf{X}\big)-F_{0}\mathbf{X}~.
\end{equation}
Since $F_{0}$ is a constant, the last term acts as a constant  driving force in the classical equation of motion.  We perform two successive unitary transformations on this Hamiltonian, so that the final transformation is given by, 
\begin{equation}
	U_{3}\big(\mathbf{X},\mathbf{P}\big)=U_{2}(\mathbf{P})U_{1}(\mathbf{X})=\exp\bigg[-i\frac{F_{0}}{A_{1}}\mathbf{P}\bigg]\exp\bigg[i\frac{C_{0}}{4B_{0}}\mathbf{X}^{2}\bigg]~~~~\text{with}~~~A_{1}=A_{0}-\frac{C_{0}^{2}}{4B_{0}}~.
\end{equation}
The first transformation is just the one that appears in Eq. (\ref{U1}). The final form of the Hamiltonian, after these transformations  are   performed,
can be written in the following simplified form
\begin{equation}\label{HOwithF0}
	\mathbf{H}_{HO}\big(\mathbf{X},\mathbf{P}\big)=U_{3}\big(\mathbf{X},\mathbf{P}\big)\mathbf{H}_{F}\big(\mathbf{X},\mathbf{P}\big)U_{3}\big(\mathbf{X},\mathbf{P}\big)^{\dagger}=\frac{A_{1}}{2}\mathbf{X}^{2}+\frac{B_{1}}{2}\mathbf{P}^{2}-\frac{F_{0}^{2}}{2A_{1}}\mathbf{I}~,
\end{equation}
which, as the subscript above suggests, is just that of a HO with a constant term added (here $B_1=B_0$). Notice that this constant term actually shifts the energy eigenvalues of the system,  so that the analogues of the relations in Eq. (\ref{regularization}) and Eq. (\ref{mean}) can be written without 
introducing the constant $a$ by hand, and are given by, 
\begin{equation}
	\text{Tr}\bigg[\mathbf{H}_{HO}-\frac{F_{0}^{2}}{2A_{1}}\mathbf{I}\bigg]=0~~~\rightarrow~~~\zeta\bigg(-1,\frac{1}{2}-\frac{F_{0}^{2}}{2A_{1}}\bigg)=0 ~~~\text{i.e.}~~~F_{0}^{2}=\frac{A_{1}\omega}{\sqrt{3}}~=\frac{B_{0}}{\sqrt{3}}\bigg(A_{0}-\frac{C_{0}^{2}}{4B_{0}}\bigg)^{2}~.
\end{equation}
The cost evaluated using this $F_{0}$ is just the same as in Eq. (\ref{GHOcost}) for generalised oscillator without the driving force.

This example actually indicates an alternative interpretation of the trace regularization condition (\ref{regularization}) 
on the  displaced Hamiltonian $\tilde{\mathbf{H}}=\mathbf{H}-a\mathbf{I}$, namely,  it is not always necessary  to introduce an additional  
term $-a\mathbf{I}$ in the original Hamiltonian, since the displaced one ($\tilde{\mathbf{H}}$)
can be thought of as that of the Hamiltonian of an ordinary HO with shifted energy eigenvalues, which, in turn, can be obtained through successive unitary transformations from that of a generalised oscillator with an extra linear potential. The  role played by the mean value $a$ is the same as that of 
the constant driving force $F_{0}$, and this can be chosen in such a way  that both methods give identical values of the bi-invariant
costs. Obviously, as can be seen by 
closely inspecting the Hamiltonian in Eq. (\ref{HOwithF0}),  the constant  force term $F_0$ does not change the quantitative
value of the cost, rather its role  is to provide an `in-build' regulator of the norm. 

The important role played by the unitary invariance of the cost functions in these arguments is to be noticed. Since the cost in Eq. (\ref{p=1cost})  is unitary invariant, the cost of preparing the  operators generated by the Hamiltonian of generalised oscillator in the presence of a constant driving force is the same as that of an HO with shifted energy values.

\subsection{ Complexity geometry generated by   the isotonic oscillator Hamiltonian}
We have seen  in the previous section that all the coefficients present in the original Hamiltonian  may not affect the numerical 
value of the norm; rather, they play different roles in determining the cost itself.  To see a similar effect of the coefficients 
present in a quantum mechanical Hamiltonian, we consider a generalisation of the HO Hamiltonian with addition of a singular 
term, i.e. in the presence of a   potential barrier at the origin. The resulting Hamiltonian, known as the isotonic oscillator, is written as 
\begin{equation}\label{Iso}
	\mathbf{H}_{IO}=\frac{A_{0}}{2}\mathbf{X}^{2}+\frac{B_{0}}{2}\mathbf{P}^{2}+\frac{D_{0}}{2\mathbf{X}^{2}}~.
\end{equation}
We can add a constant term in this Hamiltonian as well and interpret it, as explained above,  as the displaced Hamiltonian,  however here we 
omit it for convenience.  
The energy eigenvalues and the solutions of the corresponding Schrodinger equation for this Hamiltonian are well studied in the literature - see, e.g.,  \cite{YJ,CPRS,IS}. For our purposes in this section we only need the energy eigenvalues,  these are given by
\begin{equation}
	E_{nn}=\big(2n+1+\epsilon\big)\omega_{is}~,~~~\text{where}~~\omega_{is}=\sqrt{A_{0}B_{0}}~,~~\text{and}~~~\epsilon=\frac{1}{2}\sqrt{1+\frac{4D_{0}}{B_{0}}}~.
\end{equation}
Here  $n=0,1,2.\cdots.$ can take only positive integer values.  
The energy eigenvalues are equidistant, just like those of HO, but it does depend on the strength of the singular potential term.  Following the analogous procedure as that of the HO, we arrive at the $\zeta$ function associated with the displaced Hamiltonian $\tilde{\mathbf{H}}_{IO}=\mathbf{H}_{IO}-a\mathbf{I}$ (see Eq. (\ref{zetaHdis})) 
\begin{equation}
	\zeta_{\tilde{\mathbf{H}}_{IO}}(k)=\sum_{n=0}^{\infty}\frac{1}{\big(E_{nn}-a\big)^{k}}~~=\Big(\frac{\omega_{is}}{2}\Big)^{-k}\sum_{n=0}^{\infty}\bigg( \Big(n+\frac{1}{2}\big(\epsilon+1\big)\Big)-\frac{2a}{\omega_{is}}\bigg)^{-k}~=\Big(\frac{\omega_{is}}{2}\Big)^{-k}\zeta\bigg(k,\frac{1}{2}\big(\epsilon+1\big)-\frac{2a}{\omega_{is}}\bigg)~.
\end{equation}
As before,  the final expression is obtained through the definition of the Hurwitz $\zeta$ function. The criterion 
in Eq. (\ref{regularization}) now implies that the mean value $a$ is 
\begin{equation}
	\zeta\bigg(-1,\frac{1}{2}\big(\epsilon+1\big)-\frac{2a}{\omega_{is}}\bigg)=0~~~\rightarrow~~a=\frac{\omega_{is}}{2}\bigg(\frac{\epsilon}{2}\pm\frac{1}{2\sqrt{3}}\bigg)~.
\end{equation}
The norm of the Hamiltonian,  in turn,  is given by
\begin{equation}
	F^{2}_{r}\big(\mathbf{H}_{IO}\big)=\lambda_{0}^{2}\text{Tr}_{r}\Big[\Big(\mathbf{H}_{IO}-a\mathbf{I}\Big)^{2}\Big]=\lambda_{0}^{2}\zeta_{\tilde{\mathbf{H}}_{IO}}(-2)=\lambda_{0}^{2}\Big(\frac{\omega_{is}}{2}\Big)^{2}\zeta\Big(-2,\frac{1}{2}\mp\frac{1}{2\sqrt{3}}\Big)~.
\end{equation}
Since the norm should be always positive,  we have the final expression  for it to be 
\begin{equation}
	g_{IJ}Y^{I}Y^{J}=F^{2}_{r}\big(\mathbf{H}_{IO}\big)=\frac{\sqrt{3}\lambda_{0}^{2}}{108}\Big(\frac{\omega_{is}}{2}\Big)^{2}~~=\lambda_{1}^{2}\omega^{2}_{is}.
\end{equation}
Though the expression for the norm in this case is formally similar to that of HO (the only difference between the second expression of 
\eqref{GHOcost} and the above cost is that  the frequency of the HO is replaced by $\omega_{is}/2$ 
for an isotonic oscillator),  it is important to notice the role played by the strength of the singular term $D_{0}$ in determining the mean value $a$ - unlike the previous case depending on the value of $\epsilon$ and hence $D_{0}$, the mean $a=\frac{\omega_{is}}{4}\Big(\epsilon-\frac{1}{\sqrt{3}}\Big)$ can be positive as well as negative.


\section{Complexity geometry associated with the Caldirola–Kanai Hamiltonian }\label{ckosc}
As an illustrative example of the  general  result derived above, we now find out the CG  generated by the damped Caldirola–Kanai (CK) oscillator \cite{PC,EK}. For this  Hamiltonian the coefficients in Eq. (\ref{H1t}) are given by
\begin{equation}
	A_{1}(t)=M\omega^{2}\exp\big[\Delta t\big]~,~~B_{1}(t)=\frac{1}{M}\exp\big[-\Delta t\big]~,
\end{equation} 
where $M,\omega,\Delta$ are three positive constants. In the limit of $\Delta\rightarrow0$, this Hamiltonian reduces to that of a  HO of mass $M$ and frequency $\omega$. With these coefficients,  a solution of the auxiliary equation Eq. (\ref{auxiliary2})  can be written as \cite{CK}
\begin{equation}
	\rho(t)=\frac{1}{\sqrt{M\omega_{0}}}\exp\big[-\Delta t/2\big]~,~~ \text{where}~~\omega_{0}^{2}=\omega^{2}-\frac{\Delta^{2}}{4}~.
\end{equation}
Using this solution of the auxiliary in  expression (\ref{NormH1final}), we see that  the norm of the CK Hamiltonian is given by 
\begin{equation}\label{NormCK}
	F^{2}_{r}\big(\mathbf{H}_{CK}\big)=4\omega^{2}\bigg(\lambda_{12}^2\frac{\omega^{2}}{\omega_{0}^{2}}-\lambda_{2}^{2}\bigg)~,~~~
	\lambda_{12}^2=\lambda_{1}^{2}+\lambda_{2}^{2}~.
\end{equation}
Therefore, even though the Hamiltonian itself is an explicit function of time,  the associated bi-invariant cost function is actually 
independent of time. This fact is, of course, not true in general.
Here, it is related to the fact that the matrix elements of the CK Hamiltonian are time-independent in the
basis of the invariant operator eigenfunctions. \footnote{This does not, however, mean that the total mechanical energy of the system when it is in  $n$th  quantum state 
	is time-independent. For such damped systems, the mechanical energy is different from the expectation of the Hamiltonian \cite{CK}.} This result makes the computation of the cost functional, i.e. the total cost along a given path on the space of unitaries rather trivial, and in this case the total cost along the path generated by the damped CK Hamiltonian is equal to $F_{r}\big(\mathbf{H}_{CK}\big)$ itself.\footnote{As in Eq. (\ref{complexitydef}), we assume that the path is parameterized in such a way that the parameter takes values $0$ and $1$, at the starting and the ending points, respectively.}  Notice that the above statement does not imply that it is  the  minimum  value of the cost of the evolution operator,  since the curve along the cost is evaluated is generated by a time-dependent generator, and hence,  it  may not be the geodesics - which is  generated by a constant generator. Comparing the norm in Eq. \eqref{NormCK} with the second expression in Eq. (\ref{GHOcost}), we also note that the cost function associated with a CK oscillator Hamiltonian is formally the same as that  of the HO of unit mass and  frequency $\omega_{eq}=\frac{\omega}{\lambda_{1}}\sqrt{\Big(\lambda_{12}^{2}\frac{\omega^{2}}{\omega_{0}^{2}}-\lambda_{2}^{2}\Big)}$.

\section{Decomposition formula for the time evolution operator of an OTMF}\label{decomp}
In this appendix we briefly show  the  derivation of  an exact form of the  time-evolution operator for an OTMF, and obtain the decomposition formulas  by using a $2 \times 2 $ matrix 
representation of the 
generators of the corresponding Lie algebra. We consider the following matrix representation of the $su(1,1)$ Lie algebra generators 
\begin{equation}
	K_0= \frac{1}{2}\left(
	\begin{array}{ccc}
		-1 &0   \\
		0 & 1 
	\end{array}
	\right)~, 
	~~ K_+= \left(
	\begin{array}{ccc}
		0 &0   \\
		1 & 0 
	\end{array}
	\right)~,~~\text{and}
	~~ K_-= \left(
	\begin{array}{ccc}
		0 & -1   \\
		0 & 0 
	\end{array}
	\right)~.
\end{equation}
As can be easily checked, these generators satisfy the commutation relations given in Eq. \eqref{su11}. Using these forms for the generators in the decomposed 
form for the time evolution operator for an OTMF  (see the second expression in Eq. \eqref{UOTMF}) we 
obtain an exact matrix  representation for it to be of the form
\begin{equation}
	\mathcal{U}(t,0)= \exp\Big[c_{+}(t)K_{+}\Big]\exp\Big[c_{0}(t)K_{0}\Big]\exp\Big[c_{-}(t)K_{-}\Big]~=
	\left(
	\begin{array}{ccc}
	e^{-\frac{c_0(t)}{2}} & -c_-(t) e^{-\frac{c_0(t)}{2}}    \\
		c_+(t)e^{-\frac{c_0(t)}{2}}  & e^{\frac{c_0(t)}{2}} - c_+ c_-e^{-\frac{c_0(t)}{2}} 
	\end{array}
	\right)~.
\end{equation}
On the other hand, from the first expression for the evolution operator in \eqref{UOTMF}, we can also obtain 
\begin{equation}
	\mathcal{U}(t,0)= \exp\Big[b_{0}(t)K_{0}+b_{+}(t)K_{+}+b_{-}(t)K_{-}\Big]~=
	\left(
	\begin{array}{ccc}
		\cos \chi - \frac{b_0}{2 \chi } \sin \chi  & -\frac{b_-}{\chi}  \sin \chi  \\
		\frac{b_+}{\chi}  \sin \chi & 	\cos \chi + \frac{b_0}{2 \chi } \sin \chi
	\end{array}
	\right)~,
\end{equation}
where $\chi=(b_+b_--\frac{b_0^2}{4})^{1/2}$~.  Comparing these two matrix representations of the time evolution operator, we obtain the relations
between two sets of time-dependent coefficients as given in Eq. \eqref{relbc}.


\end{document}